\documentclass{JHEP3}
\usepackage{amssymb}
\usepackage{amsfonts}
\usepackage{amsbsy}
\usepackage{amsmath}
\usepackage{amsthm}
\usepackage{graphicx}
\usepackage{epstopdf}
\usepackage[vcentermath]{youngtab}
\usepackage{multirow}
\usepackage{latexsym}
\usepackage{array}

\usepackage[nosort]{cite}

\def\n3a{t}

\def\tr{{\mathrm{tr}}}

\def\n{{V_A}}

\newcommand\field[1]{{\ensuremath{\mathbb{{#1}}}}}

\newcommand{\sR}{\mathcal{R}}
\newcommand{\sT}{\mathcal{T}}
\newcommand{\GG}{\mathcal{G}}

\newcommand{\sA}{\mathcal{A}}

\newcommand{\OO}{\mathcal{O}}

\def\ov{\over}

\def\lam{{\lambda}}

\def\vev#1{\langle#1\rangle}
\def\det{{\rm det}}
\def\tr{{\rm tr}}

\def\eq#1{(\ref{#1})}

\def\Xh{{\hat{X}}}
\def\Sh{{\hat{S}}}

\def\Om{{\Omega}}

\def \lam {\lambda}

\def \ra {\rightarrow}

\def\BB{{\cal B}}

\def\Sh{{\hat{S}}}
\def\sh{{\hat{s}}}

\def\fh{{\hat{f}}}

\def\tx{{\tilde{x}}}
\def\ty{{\tilde{y}}}

\def\pls{{[+]}}
\def\Neron{{N\'eron}}

\newcommand{\be}{\begin{equation}}
\newcommand{\ee}{\end{equation}}
\newcommand{\bea}{\begin{eqnarray}}
\newcommand{\eea}{\end{eqnarray}}

\newcommand{\bln}{\begin{align}}
\newcommand{\eln}{\end{align}}
\newcommand{\bst}{\begin{split}}
\newcommand{\est}{\end{split}}
\newcommand{\bi}{\begin{itemize}}
\newcommand{\ei}{\end{itemize}}
\newcommand{\ben}{\begin{enumerate}}
\newcommand{\een}{\end{enumerate}}

\newcommand{\bpm}{\begin{pmatrix}}
\newcommand{\epm}{\end{pmatrix}}

\title{F-Theory and the Mordell-Weil Group\\
of Elliptically-Fibered Calabi-Yau Threefolds}

\author{David R. Morrison$^{1}$ and Daniel S. Park$^2$\\
\\
$^1$Departments of Mathematics and Physics\\
University of California, Santa Barbara\\
Santa Barbara, CA 93106, USA\\
\\
$^2$Center for Theoretical Physics\\
Department of Physics\\
Massachusetts Institute of Technology\\
Cambridge, MA 02139, USA\\
\\
\\
{\tt drm} {\rm at} {\tt math.ucsb.edu},
{\tt dspark81} {\rm at} {\tt mit.edu}
}

\preprint{UCSB Math 2012-27, MIT-CTP-4388}

\abstract{The Mordell-Weil group of an elliptically fibered Calabi-Yau
threefold $X$ contains information about the abelian sector of
the six-dimensional theory obtained by compactifying
F-theory on $X$. After examining features of the
abelian anomaly coefficient matrix and $U(1)$
charge quantization conditions of general F-theory vacua,
we study Calabi-Yau threefolds
with Mordell-Weil rank-one as a first step towards understanding the
features of the Mordell-Weil group of threefolds in more detail.
In particular, we generate an interesting class of F-theory
models with $U(1)$ gauge symmetry that have matter
with both charges $1$ and $2$.
The anomaly equations --- which relate the N\'eron-Tate
height of a section to intersection numbers between
the section and fibral rational curves of the manifold ---
serve as an important tool in our analysis.}

\begin{document}

\section{Introduction and Summary}

The abelian sector of F-theory backgrounds
is interesting from at least two different points of view.
From the point of view of F-theory phenomenology,
understanding how to protect and break various
$U(1)$'s in F-theory model building is essential
to constructing models with desired
properties.\footnote{The literature on $U(1)$'s
in F-theory model building is quite vast; recent works include
\cite{GrimmWeigand,DudasPalti,Marsano,DMSS,MSS,GKPW}.}
Meanwhile, from the point of view of addressing the
question of 6D string universality \cite{KTUniv},
a systematic understanding of what one could get
in string theory --- especially F-theory --- is crucial.
Such an understanding of the abelian sector
of F-theory has yet to be gained.
In this note, we approach abelian gauge
symmetry from the latter viewpoint.

The six-dimensional string universality conjecture
is the conjecture that all ``consistent" six-dimensional supergravity
theories with minimal supersymmetry\footnote{By denoting
a supergravity theory ``six-dimensional," we are further assuming
that the theory has flat six-dimensional Minkowski space
as a stable solution of the theory.} are embeddable in string theory
\cite{KTUniv}.
Much progress has been made on verifying this conjecture
by focusing on theories with non-abelian gauge symmetry
\cite{KTBound,KMT1,KMT2,KPT}.
Much more, however, needs to be understood upon
introducing abelian gauge symmetry to the picture
\cite{PTaylor}.

In particular, it is not well understood
what kind of $U(1)$ charges are allowed
in F-theory.
A simple version of the problem
is to ask what kind of charges are allowed
for the matter of a six-dimensional
F-theory background whose gauge group
is given by $\GG =U(1)$.
We currently do not know the answer even
to this seemingly innocent problem.
In this note,
we take some first steps towards improving
the current status.

The abelian sector of six-dimensional
F-theory backgrounds --- obtained
by compactification on an elliptically
fibered Calabi-Yau threefold $X$
\cite{MV1,MV2} ---
contains information about the Mordell-Weil
group of the threefold $X$.
In particular, the rank of the abelian gauge
group is equal to the Mordell-Weil rank \cite{MV2},
while the anomaly coefficient matrix
of the abelian gauge fields turns out to be
the \Neron-Tate height pairing matrix
of the Mordell-Weil generators \cite{anomint}.
Therefore, in order to understand
the abelian sector of supersymmetric
F-theory backgrounds,
one must study the Mordell-Weil
group of elliptically fibered Calabi-Yau manifolds.\footnote{
We note that the Mordell-Weil group has been studied
in various contexts in string theory. The Mordell-Weil group
of elliptically fibered surfaces has been studied using
string junctions \cite{Gaberdiel:1997ud,DeWolfe:1998zf}
in \cite{Fukae:1999zs,Guralnik:2001jh}.
The torsion subgroup of the Mordell-Weil group
has been studied for elliptically fibered threefolds
in \cite{Aspinwall:1998xj}.
It is also possible to study the Mordell-Weil group of
$T^4$ fibered manifolds --- this has been done for
certain $T^4$ fibered Calabi-Yau
threefolds in \cite{Donagi:2008ht}.}

The Mordell-Weil group of an elliptic fibration
--- which is the group of rational sections of the fibration ---
is a rather elusive mathematical object to study.
We make an initial step in this note to
understand the Mordell-Weil group of an elliptically
fibered Calabi-Yau threefold from the
point of view of F-theory.\footnote{The Mordell-Weil
group of elliptically fibered threefolds is a subject
of interest also in pure mathematics; some recent
works on this subject are
\cite{HulekKloosterman,Alexander,rankbound,ZariskiTriple}.}
In particular, we focus on a very
simple class of manifolds ---
namely, Calabi-Yau threefolds fibered over $\field{P}^2$
with no enhanced gauge symmetry and
Mordell-Weil rank-one.
Compactification of F-theory on such a manifold
yields a six-dimensional supergravity theory
with no tensor multiplets and gauge group $U(1)$.
We concern ourselves with understanding
what kind of charges of matter are allowed
for such theories.

Strong constraints on the non-abelian sector of
six-dimensional F-theory backgrounds are imposed
by the Kodaira condition
\cite{Kodaira,Kawamata,Fujita,Nakayama,GrassiKod}. 
The Kodaira condition ``bounds" the anomaly coefficients
associated to each non-abelian gauge group,
which in turn restricts the matter representations
charged under the non-abelian gauge groups.
Similar bounds on the height-pairing matrix
of the Mordell-Weil generators, if they exist,
would lead to constraints
on charges of the abelian sector.

Let us summarize the main results of this note.
\ben
\item We explicitly compute the height pairing matrix
of a given set of Mordell-Weil generators and
discuss their properties for general Calabi-Yau
threefolds.
\begin{itemize}
 \item In particular, the self-height pairing
 of the Mordell-Weil generator $\sh$ of a
 Calabi-Yau threefold fibered over $\field{P}^2$ with
 Mordell-Weil rank-one is parameterized by
 a single non-negative integer $n$,
 in the absence of enhanced
 non-abelian gauge symmetry.
 A bound on this number $n$
 would serve as an analogue of the
 Kodaira bound for these 
 theories.\footnote{It is worth pointing out that only a finite number
of values of $n$ could possibly occur among elliptic Calabi-Yau
threefolds with Mordell-Weil rank-one, although we have no way to
calculate the maximum value.  This is because it is known that
elliptic Calabi-Yau threefolds form only finitely many
algebraic families \cite{GrassiKod,alg-geom/9305003,MR1272978},
and in each family with Mordell-Weil rank-one, the value of $n$
is constant.}
\end{itemize}
\item Using anomalies,
we show that when one assumes that the charge
of the matter is either $1$ or $2$ there are
only nine distinct possible theories each with $n=0,\cdots,8$.
\item We explicitly construct seven of these nine
theories, namely theories with $n=0,\cdots,6$.
\een

The structure of this note is as the following.
We first review six-dimensional F-theory backgrounds in
section \ref{s:6Dreview}.
We then compute the abelian anomaly coefficients of
an F-theory background after reviewing how to extract information
of the abelian sector of F-theory models from the geometry
in section \ref{s:ab}, {\it i.e.,} we arrive at result (1).
In particular, we show how the analysis simplifies in the
case of pure abelian theories with no tensor multiplets.
In section \ref{s:anom}, we show how the charge is restricted
in the case of pure abelian models with no tensor multiplets
when the anomaly coefficient is given. In particular,
we reach result (2) in this section.
We construct specific models with Mordell-Weil rank
$1$ in more detail in section \ref{s:rank1}, {\it i.e.,}
we arrive at result (3).
We sketch questions and future directions in
section \ref{s:directions}.

\section{Review of Six-Dimensional Supergravity Theories} \label{s:6Dreview}

We review relevant facts about six-dimensional $(1,0)$
supergravity theories and F-theory backgrounds in this section.
We also explain why the Kodaira condition restricts the
charged matter structure for the non-abelian sector
of F-theory models briefly. The presentation of this section
is rather condensed. Further details can be found in
\cite{PTaylor,anomint}.

\begin{table}[t!]
\center
  \begin{tabular}{ | c | c |}
  \hline
  Multiplet & Field Content\\ \hline
  Gravity & $(g_{\mu \nu}, \psi^+_\mu, B^+_{\mu \nu}) $    \\  \hline
  Tensor & $(\phi, \chi^-, B^-_{\mu \nu})$     \\  \hline
  Vector & $(A_{\mu}, \lam^+)$     \\  \hline
  Hyper & $(4\varphi, \psi^-)$    \\  \hline
    \end{tabular}
  \caption{Six-dimensional (1,0) supersymmetry multiplets. The signs on
   the fermions indicate the chirality. The signs on antisymmetric tensors
   indicate self-duality/anti-self-duality.}
\label{t:mult}
\end{table}

The low-energy data of six-dimensional supergravity theories
can be parameterized by its massless spectrum $S$,
anomaly coefficients $\{ b \}$ and a ``modulus'' $j$.
The massless particles come in BPS multiplets of the
$(1,0)$ supersymmetry algebra given in table \ref{t:mult}.
The massless spectrum $S$ is specified by the number
of tensor multiplets $T$, the (global) gauge group
\be
\GG = \prod_{\kappa=1}^n \GG_\kappa \times \prod_{i=1}^\n U(1)_i
\ee
--- where $\GG_\kappa$ are simple non-abelian gauge group
factors --- and the matter representation of the hypermultiplets.
$\n$ denotes the number of abelian gauge group factors.
There is one gravity multiplet in the theory.

The anomaly coefficients $\{ b \}$ are a set of $SO(1,T)$ vectors.
To each non-abelian gauge group factor $\GG_\kappa$, there is an
$SO(1,T)$ vector $b_\kappa$ associated to it, which we call the
``anomaly coefficient of $\GG_\kappa$."
For the abelian sector, there is an associated
$\n \times \n$ matrix $b_{ij}$ whose components are also $SO(1,T)$
vectors. We call this matrix the ``abelian anomaly coefficient matrix."
Abelian vector fields are defined only up to linear transformations.
Imposing charge minimality constraints on the abelian gauge symmetry,
the abelian gauge fields are defined up to $SL(\n,\field{Z})$ ---
it follows that $b_{ij}$ transforms as a bilinear under this $SL(\n,\field{Z})$.
There also exists a gravitational anomaly coefficient $a$ that is
needed to cancel the gravitational/mixed anomalies of the theory.

These six-dimensional theories must satisfy generalized
Green-Schwarz anomaly cancellation conditions
\cite{GS,GSW,Sagnotti:1992qw,Sadov,Riccioni1,Riccioni2}.
These anomaly cancellation conditions come from
imposing that the eight-dimensional
anomaly polynomial --- computed by adding the contribution of
all the chiral fields of the theory \cite{A,BJ,ganomalies} ---
should factor into the form
\be
I_8 \propto ({1 \ov 2} a \tr R^2 + \sum_{\kappa} {2 b_\kappa \ov \lambda_\kappa} \tr F_\kappa^2+
 \sum_{i,j} 2 b_{ij}  F_i F_j)^2
\ee
where the norm is taken with respect to an $SO(1,T)$
metric $\Om_{\alpha\beta}$. $\lambda_\kappa$ is the Dynkin
index of the fundamental representation of $\GG_\kappa$.
The explicit form of the anomaly equations have been written out, for example,
in \cite{KMT2,PTaylor,anomint,Erler,Honecker} and we do not reproduce them here.

The anomaly coefficients, along with the modulus $j$ determine
important terms in the low-energy effective Lagrangian.
The modulus $j$ is a unit $SO(1,T)$ vector parametrizing
the vacuum expectation value of the $T$ scalar fields
in the tensor multiplets.
In particular, they determine the kinetic terms for the gauge fields
\be
\propto \sum_{\kappa} ({j \cdot b_\kappa \ov \lambda_\kappa} ) \tr F_\kappa \wedge *F_\kappa
+ \sum_{k,l}  ({j \cdot b_{kl} }) F_k \wedge *F_l \,,
\ee
and the Green-Schwarz term
\be
\propto B \cdot ({1 \ov 2} a \tr R^2 +
\sum_{\kappa} {2 b_\kappa \ov \lambda_\kappa} \tr F_\kappa^2+
\sum_{i,j} 2 b_{ij}  F_i F_j) \,,
\ee
where the self-dual and anti-self-dual tensors are organized
into an $SO(1,T)$ vector and the inner-product is taken with
respect to $\Om_{\alpha\beta}$.

A six-dimensional $(1,0)$ supergravity theory can be obtained
by compactifying F-theory on an elliptically fibered Calabi-Yau
threefold $\Xh$ with a section\footnote{We use $Z$ to denote this 
``zero section" of the fibration throughout this note.}.
The anomaly coefficients have a
nice interpretation in terms of the geometry of $\Xh$.
In particular, the $SO(1,T)$ lattice on which the anomaly coefficients
live is the $H_2$ homology lattice of the base $\BB$ of the fibration.
The base $\BB$ of the fibration must be a rational surface
with $h^{1,1}(\BB)=T+1$.
The anomaly coefficients turn out to be curves --- {\it i.e.,} divisors ---
on the base which are represented by vectors in the homology lattice
$H_2 (\BB)$. In particular, the gravitational anomaly coefficient
$a$ corresponds to the canonical divisor while the coefficient
$b_\kappa$ corresponds to the degeneration locus of the fiber that yields the
$\GG_\kappa$ gauge symmetry \cite{KMT1,KMT2,MV1,MV2,BIKMSV}.

The abelian anomaly coefficients can be obtained in the following way.
Each abelian gauge field corresponds to a generator of the
Mordell-Weil group of the elliptic fibration. The
Mordell-Weil group is the group of rational sections of the fibration,
and is generated by a finite basis. The basis is defined
up to $SL(\n,\field{Z})$, which is precisely the group of redefinitions
on $U(1)$'s preserving charge minimality.
Denoting the the basis elements
of the Mordell-Weil group $\{ \sh_1, \cdots, \sh_\n \}$,
the abelian anomaly coefficients are given by
\be
b_{ij} = -\pi(\sigma(\sh_i)\cdot \sigma(\sh_j) ) \,.
\ee
Some explanation of notation is due.
$\sigma$ is a map from the Mordell-Weil group
to $H_4(\Xh)$ defined by Shioda in \cite{Shioda1},
which we refer to as the Shioda map. We explain
this map in more detail in the subsequent section,
but mention here that it is a homomorphism from the
Mordell-Weil group to the homology group $H_4$.
In other words, the group action --- or addition ---
of sections carry over to into addition
of the homology class of the corresponding sections
under the Shioda map \cite{Shioda1,Shioda2}.
The dot product is the intersection product in
the manifold $X$.
$\pi(C)$ is defined as the projection of a
curve $C$ to the base $\BB$. At the operational level,
if we denote by $B_\alpha$ the pullback of the generators $H_\alpha$
of the base homology by the projection map,
$\pi$ is defined by
\be
\pi(C) = (C \cdot B_\alpha) H^{\alpha} \,.
\ee
Here the $\alpha$ indices are vector indices of
the base homology lattice.
They are raised and lowered by the metric
\be
\Om_{\alpha \beta} = H_\alpha \cdot H_\beta \,,
\ee
where the intersection product is taken within the base
manifold in this equation.

In order for the fibration to be Calabi-Yau, the Kodaira equality
\be
-12a = \sum_\kappa \nu_\kappa b_\kappa +Y
\ee
must hold for some effective divisor $Y$,
where $\nu_\kappa$ are coefficients associated to
the reducible fiber at $b_\kappa$.
Numerical values for $\nu_\kappa$ can be found in,
for example, \cite{MV1}.
The Kodaira equality can also be thought of as a bound
\be
-12a \geq \sum_\kappa \nu_\kappa b_\kappa \,,
\ee
{\it i.e.,} it bounds the anomaly coefficients in a given base
above by $-12a$.
When there are hypermultiplets of representations with
large charges\footnote{By representations of ``large charge"
we mean representations whose components
have large charges under the
Cartan generators of the Lie group.}
under the gauge group $\GG_\kappa$,
the values $b_\kappa \cdot b_\kappa$,
$b_\kappa \cdot b_\lambda$ and $a \cdot b_\kappa$
become large. The Kodaira bound therefore restricts the
representations allowed in F-theory models by bounding $b_\kappa$.
Such restrictions placed by the Kodaira bound have been
explicitly demonstrated in the case of $T=0$ models in
\cite{KPT}.

Meanwhile, an analogous constraint on the abelian
anomaly coefficient has yet to be found. Such a constraint
would also restrict the allowed charge of matter in F-theory
models. For example, by anomaly cancellation,
\be
b_{ii} \cdot b_{ii} = {1 \ov 3} \sum_{q} x_{q} q^4 
\ee
where $x_q$ is the number of hypermultiplets with
charge $q$ under $U(1)_i$.
A bound on $b_{ii}$ would restrict the charges hypermultiplets
can have, given that there is a unit of quantization of the
$U(1)$ charges.
We explore such restrictions further in section \ref{s:anom}.

\section{The Abelian Sector of F-theory Vacua} \label{s:ab}

In this section, we review how to extract data on
the abelian sector of F-theory vacua and examine the
special case of $T=0$ backgrounds with gauge group $U(1)$
in detail.
In section \ref{ss:gen},
we give a general overview of the Mordell-Weil group
of an elliptically fibered Calabi-Yau threefold
and its relation to the abelian sector of the corresponding
six-dimensional F-theory background.
We specialize to F-theory backgrounds with
$T=0$ and gauge group $U(1)$
in section \ref{ss:teq0ab}.

\subsection{The Mordell-Weil Group and the Abelian Sector of F-theory}
\label{ss:gen}

In this section, we review relevant facts about the Mordell-Weil
group of elliptic fibrations and its relation to the abelian
sector of F-theory backgrounds.
A more thorough description of the process of extracting the physical
data of six-dimensional F-theory backgrounds
can be found in \cite{anomint,BonettiGrimm}.
More information on elliptic curves and the Mordell-Weil group
can be found in standard introductory texts on the subject,
such as \cite{Silverman}.

Six-dimensional F-theory compactifications are defined for elliptically
fibered Calabi-Yau threefolds with a section.
Let us denote such a smooth elliptic fibration by $\Xh$
and its base manifold by $\BB$.
Such manifolds have a Weierstrass representation
\be
X ~: \quad y^2 = x^3 +f xz^4 + gz^6 \,.
\ee
$f$ and $g$ are holomorphic sections of the line bundles
$-4K$ and $-6K$ respectively,
where $K$ is the canonical class of the base
manifold.\footnote{$X$ is an algebraic variety birationally equivalent
to $\Xh$ which in general is singular. $\Xh$ can
be obtained from $X$ by blowing up its singularities.
Although $X$ may have more than one Calabi-Yau resolution,
the physics in the F-theory limit is independent of the choice.
In certain cases, some aspects of the matter representation
--- which comes from
codimension-two singularities in the base ---
may be ambiguous (for example, the sign of the charges of certain
fields may depend on the resolution) but the overall matter representation
is free from ambiguity.
More discussion on resolving codimension-two
singularities of elliptically fibered manifolds
can be found in \cite{Morrison:2011mb,Katz:2011qp,
Esole:2011sm,Krause:2011xj,GrimmHayashi}.}
Hence the Calabi-Yau threefold can be thought of
as an elliptic curve over a function field.
Given an elliptic curve over a field with a choice
of a ``zero point" there is an abelian group operation ---
which we denote by ``$\pls$" ---
on the points of the elliptic curve.
This operation is defined in  appendix \ref{ap:add}.

The rational sections of the fibration can be thought of
as rational points on the elliptic curve over the function field
of the base $\BB$.
As elaborated in appendix \ref{ap:add},
these points form an abelian group under $\pls$.
This group is called the Mordell-Weil group of the
elliptic curve.

The Mordell-Weil theorem states that this group
is finitely generated.  (Unfortunately, the proof does not provide
an algorithm for computing it!)
Any finitely generated abelian group can be written
in the form
\be
\field{Z} \oplus \cdots \oplus \field{Z} \oplus \GG
\ee
where $\GG$ is the torsion subgroup.
Let us denote by $\{ \sh_1, \cdots, \sh_\n \}$
a set of rational sections that generate
the non-torsion part of the Mordell-Weil group,
{\it i.e.,} a ``basis" of the
Mordell-Weil group. As mentioned in the previous
section, this basis is defined up to linear
redefinitions $SL(\n,\field{Z})$.
$\n$ is called the Mordell-Weil rank.

Let us denote the homology classes of
the sections $\{ \sh_1, \cdots, \sh_\n \}$
--- which are four-cycles within the threefold ---
by $\{ \Sh_1, \cdots, \Sh_\n \}$.
In fact, we frequently denote a rational section
by a hatted lower-case roman letter and its homology
class by the corresponding hatted upper-case letter
throughout this note.
The Shioda-Tate-Wazir theorem \cite{Wazir}
states that $\{ \Sh_1, \cdots, \Sh_\n \}$
along with the zero section $Z$,
the vertical divisors $B_\alpha$
and the ``fibral divisors" 
$T_{\kappa,I}$ generate the homology
group $H_4(\Xh)$.
The fibral divisors of $\Xh$ are topologically
rational curves fibered over codimension-one
loci $b_\kappa$ of the base.

The structure of the fibral divisors determine the
non-abelian gauge group and anomaly coefficients.
Each non-abelian gauge group $\GG_\kappa$
is associated to a curve/divisor $b_\kappa$
in the base manifold where the fiber becomes degenerate.
The irreducible components of the degenerate
fibers are generated by a set of rational curves
$\alpha_{\kappa,I}$ that correspond to the simple
roots of the Lie algebra $\GG_\kappa$.
Any rational curve that is a fiber component
along $b_\kappa$ can be written as a linear
combination of $\alpha_{\kappa,I}$ --- these rational
curves correspond to the positive roots of the
Lie algebra $\GG_\kappa$.
The fibral divisors $T_{\kappa,I}$ are obtained by
fibering these rational curves $\alpha_{\kappa,I}$
over the locus $b_\kappa$.
The monodromy invariant fiber of $T_{\kappa,I}$,
which we denote by $\gamma_{\kappa,I}$ ---
can consist of multiple rational curves.

Let us digress briefly to describe degenerate fibers
of the manifold. The rational curve components
of degenerate fibers shrink in the ``F-theory limit"
and contribute massless vector and hypermultiplets
to the six-dimensional spectrum.
This is equivalent to saying that these rational
curves satisfy
\be
c \cdot Z = c \cdot B_\alpha =0 \,.
\ee
We call the fibral rational curves --- rational curves
that are components of a degenerate fiber --- that
are fibered over some curve in the base $b_\kappa$,
``fibered" (fibral) rational curves.
Each fibered rational curve contributes two
vector multiplets and $2g$ hypermultiplets
to the massless spectrum, where $g$ is the
genus of the curve the rational curve is
fibered over.
There can be other fibral rational curves
that are isolated at codimension-two loci in the base.
We call these ``isolated" (fibral) rational curves.
Each of these curves contribute a hypermultiplet
to the six-dimensional spectrum.
For obvious reasons, we use the term ``fibral rational
curves" and ``shrinking rational curves" interchangeably.

Now we are in a position to define the Shioda map $\sigma$.
For a rational section $\sh$, let $\Sh \in H_4 (\Xh)$ be its homology class.
Then
\be
\sigma(\sh)=\Sh-Z-(\Sh \cdot Z \cdot B^\alpha -K^\alpha) B_\alpha
+ \sum_{I,J,\kappa}(\Sh\cdot \alpha_{\kappa,I}) (C_\kappa^{-1})_{IJ} T_{\kappa,J} \,.
\ee
Here $K^\alpha$ are the coordinates of the canonical
class of the base, while $C_\kappa$ is the Cartan matrix of the
Lie algebra $\GG_\kappa$ defined by
\be
(C_\kappa)_{IJ} = {2 \vev{\alpha_I,\alpha_J} \ov \vev{\alpha_I, \alpha_I}}
\ee
where $\alpha_I$ are the simple roots of $\GG_\kappa$.
There is a one-to-one
correspondence between the abelian vector fields
$\{ A_1, \cdots, A_\n \}$ and the
four-cycles $\{ \sigma(\sh_1), \cdots, \sigma(\sh_\n) \}$.
For convenience we say that $A_i$ is ``dual to"
$\sigma(\sh_i)$.\footnote{This duality is physical
in the following sense. Each six-dimensional
abelian vector field, when KK-reduced to five
dimensions, is still an abelian vector field.
Each vector field of the five-dimensional theory,
due to M-theory/F-theory duality, is obtained by
KK-reducing the M-theory three-form along a
harmonic two-form in the manifold $\Xh$.
The $A_i$ field when KK-reduced is obtained
in the M-theory dual by KK-reducing the 11D three-form
along a harmonic two-form that is Poincar\'e dual
to the four-cycle $\sigma(\sh_i)$.}
The anomaly coefficient matrix is given by
\be
b_{ij} = -\pi (\sigma(\sh_i) \cdot \sigma (\sh_j))
\ee
where $\pi$ is the projection to the
$H_2 (\BB) $ homology lattice of the base.
As we elaborate shortly, this is the generalized
N\'eron-Tate height pairing for elliptically fibered
threefolds, and $b_{ij}$
constitutes the height pairing matrix
of the elliptic fibration.

Let us briefly review the mathematical significance of
the Shioda map.
For elliptically fibered Calabi-Yau threefolds, there is a natural
inner-product on two elements $S, S'$
of $H_4 (\Xh,\field{Z}) \cong H^{2,2} (\Xh)$ with values in $H_2 (\BB)$
\cite{Wazir}.
It is, in fact, given by
\begin{align}
\begin{split}
\langle~ ,~  \rangle ~:~
H_4 (\Xh) \times H_4 (\Xh)
\quad
&\rightarrow 
\quad
H_2 (\BB) \\
( S, S' )
\quad
&\mapsto
\quad
- \pi( S \cdot S')
\end{split}
\end{align}
The Shioda map $\sigma$ is the map from
the Mordell-Weil group
to the orthogonal complement of the space
spanned by the zero section $Z$,
the vertical divisors $B_\alpha$
and the fibral divisors $T_{\kappa,I}$
under this inner-product.
In other words,
\be
\langle \sigma(\sh),  C \rangle = 0
\ee
for any section $\sh$ and an
element $C$ which is an element
of the subspace of $H_4 (\Xh)$
spanned by $Z$, $B_\alpha$ and $T_{I,\kappa}$.
Following \cite{Shioda2},
let us denote this subspace by ``$T$."
Using the fact that $\sigma(\sh)$ is a
projection of the homology class of $\sh$
to $H_4(X)/T$,
it can be shown that $\sigma$ is a homomorphism
from the Mordell-Weil group to the homology lattice
\cite{Shioda1,Shioda2,Wazir}, {\it i.e.,}
\be
\sigma(\sh \pls \sh') =\sigma(\sh) + \sigma(\sh') \,,
\ee
where the addition on the right-hand-side is
the addition defined for the homology group.
Then, the inner-product
\be
\langle \sigma(\sh),  \sigma(\sh') \rangle
= -\pi (\sigma(\sh) \cdot \sigma (\sh'))
\ee
is the \Neron-Tate height pairing
of rational sections of the elliptic fibration.

As mentioned previously,
each abelian vector field $A_i$
is dual to $\sigma (\sh_i)$.
Also, each element of the Cartan of the non-abelian
gauge group $\sA_{\kappa,I}$ is dual to the
fibral divisors $T_{\kappa,I}$ in the same sense.
A multiplet coming from a fibral rational curve $c$
has charge $S \cdot c$ under the Cartan vector field
dual to a four-cycle $S$.

By construction $\sigma (\sh_i) \cdot c =0$
for any fibered rational curve $c$, and therefore no vector multiplet
is charged under the abelian vector fields $A_i$, as desired.
Hence only hypermultiplets coming from isolated rational
curves are charged under the abelian vector fields.
For any isolated rational curve $c$, the hypermultiplet
corresponding to it has charge
\begin{align}
\sigma(\sh_i) \cdot c=(\Sh_i \cdot c)
+ \sum_{I,J,\kappa}(\Sh_i \cdot \alpha_{\kappa,I}) (C_\kappa^{-1})_{IJ} (T_{\kappa,J} \cdot c) \,.
\end{align}
under $A_i$.
Since the intersection numbers  $(\Sh_i \cdot c)$
and $(T_{\kappa,J} \cdot c)$ are integral, we
see that the unit charge of the abelian vector field
$A_i$ is given by the inverse of
the least common multiple of  $\{ \det (C_\kappa) \}$,
where $\kappa$ runs over the gauge groups for which
$\alpha_{I,\kappa} \cdot \Sh_i \neq 0$ for
some root $\alpha_{I,\kappa}$.
In particular, when $c \cdot \Sh_i=0$
for all fibered rational curves $c$,
the charges of the matter under the abelian vector fields
are integral.

Let us end this section by computing the
abelian anomaly coefficient matrix and examining its properties.
The anomaly coefficient matrix can be written as
\be
b_{ij} = -\pi(\Sh_i \cdot \Sh_j)-K
+(n_i + n_j) -(\sR^{-1}_\kappa)_{IJ} (\Sh_i \cdot \alpha_{I,\kappa})
(\Sh_j \cdot \alpha_{J,\kappa}) b_\kappa \,.
\label{bij}
\ee
$n_i$ is the locus along which section $S_i$
intersects the zero section, {\it i.e.,}
\be
n_i \equiv \pi(\Sh_i \cdot Z) \,.
\ee
$K$ is the canonical class of the base manifold.
The matrix $\sR_\kappa$ is the normalized root matrix
of Lie group $\GG_\kappa$
\be
(\sR_\kappa)_{IJ} = {2 \vev{\alpha_I ,\alpha_J} \ov \vev{\alpha,\alpha}_\text{max}}
\ee
where $\vev{\alpha,\alpha}_\text{max}$ is the length of
the longest root of the Lie group.
We have used the following equalities in arriving at
\eq{bij}:
\begin{align}
Z \cdot Z \cdot B_\alpha = K_\alpha, \quad
Z \cdot B_\alpha \cdot B_\beta = \Sh \cdot B_\alpha \cdot B_\beta = \Om_{\alpha\beta}, \quad
Z \cdot f = \Sh \cdot  f =1 \,.
\end{align}
The first of these equations follow from the
adjunction formula, and the Calabi-Yau condition ---
by these two facts, the canonical class of
$Z$ is given by the restriction of the divisor class $Z$
to itself, and hence
\be
Z \cdot Z \cdot B_\alpha = Z|_Z \cdot H_\alpha
=K \cdot H_\alpha = K_\alpha \,,
\ee
since $Z$ is topologically just the base manifold.
The intersection products in the second
and third terms are taken in the base manifold $Z$.

Such a relation actually holds for any section $\Sh$,
{\it i.e.,}
\be
\Sh \cdot \Sh \cdot B_\alpha = \Sh|_\Sh \cdot H_\alpha
=K \cdot H_\alpha = K_\alpha \,.
\ee
In general, $\Sh$ is topologically
a manifold obtained by blowing
up points on the base manifold $\BB$.
Its $H_\alpha$ components, however, coincide
with those of the canonical class of the base manifold.
Therefore
\be
b_{ii} = -2K+2n_i  -(\sR^{-1}_\kappa)_{IJ} (\Sh_i \cdot \alpha_{I,\kappa})
(\Sh_i \cdot \alpha_{J,\kappa}) b_\kappa \,.
\ee
$\sR_\kappa$ is clearly a positive-definite matrix and since
$b_\kappa$ is effective, it can be further seen that
\be
b_{ii} \leq -2K+2n_i \,.
\ee

\subsection{$T=0$ Theories with Gauge Group $U(1)$}
\label{ss:teq0ab}

In this section, we specialize to F-theory backgrounds with
$T=0$ and gauge group $U(1)$ and examine the
abelian sector. The abelian charges in this case turn
out to be quantized to be integers. Also, the abelian
anomaly coefficient is parametrized by
a single integer $n$.

Six-dimensional $T=0$ theories are obtained by
F-theory compactifications on an elliptically fibered
Calabi-Yau manifold over $\field{P}^2$.
As $h^{1,1} (\field{P}^2)=1$, $H^2(\BB)$ is
generated by the hyperplane class $H$. There
is only one vertical divisor, $B$, obtained by
pulling back the hyperplane class with respect
to the projection map.
As the homology lattice of the base is one-dimensional,
the anomaly coefficients of the theory --- being
vectors in this lattice --- are numbers.

If we restrict our attention to theories with gauge group $U(1)$,
the situation simplifies further.
The $H_4 (\Xh)$ lattice is generated by the zero section $Z$,
the vertical divisor $B$, and $S \equiv \sigma(\sh)$
where $\sh$ is the generator of the Mordell-Weil group.
All the fibral rational curves are isolated.
It is clear that the unit charge for the abelian
vector field is $1$ by the discussion in the last
subsection, as there are no fibral divisors.

The Shioda map simplifies to
\be
\sigma(\sh')
=\Sh' -Z-(\Sh' \cdot Z \cdot B+3) B
\ee
for any rational section $\sh'$.
$\Sh'$ is the homology class of $\sh'$.
We have used
\be
Z \cdot Z \cdot B = Z|_Z \cdot B|_Z =
K \cdot H =-3 \,,
\ee
which follows from the adjunction formula
and the fact that the canonical class of the base
is given by $K=-3H$.

The Weierstrass model for this elliptic fibration is given by
\be
X ~: \quad y^2 = x^3 +F_{12} xz^4 + G_{18} z^6
\ee
where $F_{12}$ and $G_{18}$ are holomorphic sections of
$12H$ and $18H$.\footnote{At the operational level,
this means that $F_{12}$ and $G_{18}$ are homogeneous
polynomials of degrees $12$ and $18$ with respect to the
projective coordinates of $\field{P}^2$.}
The Mordell-Weil generator $\sh$ must have the form
\be
\sh ~:~[x,y,z]=[f_{2n+6},f_{3n+9},f_n],
\ee
for mutually relatively prime polynomials $f_{2n+6}$,
$f_{3n+9}$ and $f_n$.\footnote{By ``mutually relatively prime"
we mean that there does not exist a polynomial $b$
of degree $\geq 1$ such that $b | f_n$,
$b^2 | f_{2n+6}$ and $b^3 | f_{3n+9}$.
Anytime we write a section in this projective form,
we assume that the three projective components
are ``mutually relatively prime" in this sense.}
$f_k$ are polynomials of degree $k$
with respect to the $\field{P}^2$ coordinates.
Then it is clear that the intersection of this
section with $Z$ is at $nH$, {\it i.e.,}
\be
\Sh \cdot Z \cdot B = n \,.
\ee
The Shioda map maps the section to
\be
S=\sigma(\sh) = \Sh - Z - (n+3)B \,.
\ee
The abelian anomaly coefficient matrix has
the single component
\be
b = - S \cdot S \cdot B
= 6
+2 \Sh \cdot Z \cdot B = 2(n+3) \,.
\label{b}
\ee
Therefore the anomaly coefficient is determined
by a single number $n$, which parametrizes
the degree of the curve on which $\sh$ intersects the zero-section.

As mentioned earlier, all the fibral curves are isolated.
Each isolated rational curve $c$ corresponds to a hypermultiplet
of the six-dimensional theory.
The charge of the hypermultiplet under the
abelian gauge group is given by
the intersection number
\be
c \cdot \sigma(\sh) = c \cdot \Sh \,.
\ee

As mentioned at various points in this note,
understanding the bounds on $b$ is crucial in understanding
what kind of charges are allowed in F-theory.
For F-theory backgrounds with $T=0$ and gauge group $U(1)$,
we have shown that the charges are integral and that
$b=2(n+3)$ when the generator of the Mordell-Weil group
is of the form
\be
\sh ~:~[x,y,z]=[f_{2n+6},f_{3n+9},f_n] \,.
\label{mwgenform}
\ee
Therefore the interesting question is what
the bound on the integer $n$ is.
Such a bound would play --- in $U(1)$ theories ---
the role the Kodaira bound plays in restricting non-abelian theories.

We note that given any elliptically fibered Calabi-Yau manifold
of Mordell-Weil rank $\geq 1$,
there exists a section with arbitrarily large self-height pairing,
which can be obtained by adding a given rational section many times.
Let us demonstrate this fact.
Given any section
\be
\sh' ~:~[x,y,z]=[f_{2n+6},f_{3n+9},f_n] \,,
\ee
and its homology class $\Sh'$, one can show that
the homology class $\Sh'_m$ of $m \sh'$ is given by
\be
\Sh'_m = m\Sh'-(m-1)Z-(n+3)m(m-1)B \,.
\label{sm}
\ee
To show this, one begins with the fact that
\be
\sigma(m \sh') =m\sigma(\sh')=m\Sh'-mZ-m(n+3)B
\ee
and hence
\be
\Sh'_m = m\Sh'-zZ-bB \,,
\ee
for some $z$ and $b$.
Since $m\sh'$ is also a section, its homology class $\Sh'_m$
intersects the fiber class once.
Imposing that $\Sh'_m \cdot f=1$ for the fiber class $f$,
one obtains $z=(m-1)$. Finally imposing the condition that
\be
\Sh'_m \cdot \Sh'_m \cdot B=-3\,,
\ee
one arrives at \eq{sm}.
Since
\be
\Sh'_m \cdot Z \cdot B = mn+3(m-1)-(n+3)m(m-1)= (n+3)m^2-3,
\ee
the section $m\sh'$ is of the form
\be
m \sh' ~:~[x,y,z]=[f_{2(n+3)m^2},f_{3(n+3)m^2},f_{(n+3)m^2-3}] \,,
\label{multorder}
\ee
with self-height pairing
\be
-\pi(\sigma(m\sh') \cdot \sigma(m\sh')) = 2(n+3) m^2 \,.
\ee

The relevant question to ask is what the
bound to the height of a Mordell-Weil generator is.
We do not know the answer to this question at the present.
To address this question, one must first understand 
how to discern whether a section is a generator or not.
We can do this in the case of pure $U(1)$ theories.
More precisely put, a sufficient condition for a given section to
be a Mordell-Weil generator is that the integral charges of the matter
under the dual abelian vector field is mutually
relatively prime.\footnote{We expect this to be
a necessary condition for a given section to be the generator
of the Mordell-Weil group also, based on the charge minimality
conjecture for gravity theories. Discussion of charge minimality
can be found in \cite{Polchinski,Banks:2010zn,HellermanSharpe,
SeibergTaylor}.}
In geometric terms, given a section $\sh'$,
the section $\sh'$ is a generator of the Mordell-Weil group
when the set of intersection numbers of its homology class $\Sh'$
(or equivalently $\sigma(\sh')$)
with the isolated rational curves of the manifold $c_1, \cdots, c_H$ are
mutually relatively prime.
This is because if $\sh' = m \sh$ for some $\sh$ and $|m| >1$,
$\Sh'$ is given by
\be
\Sh' = m\Sh-(m-1)Z-(n+3)m(m-1)B
\ee
where $\Sh$ is the homology class of $\sh$.
Therefore the intersection numbers of $\Sh'$ with the
isolated fibral rational curves  $c_1, \cdots, c_H$ are given by
\be
q_I \equiv \Sh' \cdot c_I = m(\Sh \cdot c_I) \,,
\ee
and hence $\{ q_I \}$ have $m$ as a common divisor.

\section{Anomaly Coefficients and Charge Constraints} \label{s:anom}

In this section, we see how the anomaly coefficient of a $U(1)$
theory with $T=0$ constrains the charges of the matter.
As seen in the previous sections, for F-theory backgrounds
with $T=0$ and gauge group $U(1)$, the abelian anomaly coefficient
matrix is given by $b=2(n+3)$ with non-negative integer $n$.
The charges of matter under the given $U(1)$ represented
by the Mordell-Weil generator
are quantized to be integers.
Hence the mixed/gauge anomaly equations are
given by
\begin{align}
18b = 36 (n+3) &= \sum_{c=1}^m c^2 n_c \label{anom1} \\
3b^2 = 12 (n+3)^2 &= \sum_{c=1}^m c^4 n_c \label{anom2}
\end{align}
where $n_c$ is the number of matter with charge $c$
and $m$ is the maximal charge.
The gravitational anomaly bound is given by
\be
\sum_{c=1}^m n_c \leq 274 \,.
\ee
It is clear from these equations that
\be
m ^2 \geq {n \ov 3} +1 \,,
\ee
{\it i.e.,} that when $n$ is large, there must be
some matter with charge at least $\sqrt{(n+3)/3}$.
The converse, obviously, does not hold.
For example, one could have theories with $n <9$
that have matter with charge $3$.
It is possible to show, however, that when $b=6$,
the only possible choice of matter content is
that there are $108$ hypermultiplets of charge $1$.

It is interesting to classify all the possible
theories with only charges $1$ and $2$.
The theories are characterized uniquely by $n$.
This is because the anomaly equations
\begin{align}
36 (n+3) &= n_1 + 4 n_2 \\
12 (n+3)^2 &= n_1 + 16 n_2
\end{align}
are solved by
\be
(n_1,n_2)=(4(n+3)(9-n),n(n+3)) \,.
\ee
It is clear that $n \leq 9$ and hence $b \leq 24$.
Let us denote these theories by $\sT_n$.

The $n=9$ ($b=24 = 2^2 \cdot 6$)
theory $\sT_9$ has $108$ hypermultiplets of charge $2$
and no hypermultiplets with charge $1$.
The correct way to describe this theory
is to treat it as a theory with $108$ hypermultiplets
of charge $1$ with $b=6$. In other words,
this theory is actually equivalent to the $n=0$
theory, if one chooses the correctly normalized
basis for the $U(1)$ vector field.
We therefore see that the range of values
allowed for $n$ is given by $n=0,\cdots,8$
when we allow only for charges $1$ and $2$
in the theory.
None of these theories violate
the gravitational anomaly bound.
We construct the theories $\sT_n$ for
$n = 0,\cdots,6$ explicitly in the subsequent section.

\section{$U(1)$ Theories with Charges $1$ and $2$} \label{s:rank1}

In this section, we construct Calabi-Yau threefolds fibered over $\field{P}^2$
that have gauge group $U(1)$ and matter with charges $1$ and $2$.
We are able to construct the examples $\sT_n$
with $n=0,1,\cdots,6$ presented in the previous section.
We have, however, not yet been able to construct the other
two models with $n = 7$ and $8$.

We first study the rational sections generated by the integral Mordell-Weil
section for $\sT_0$ and understand their properties in section
\ref{ss:beq6}. In particular, we observe how a section could
intersect a rational curve multiple times by examining the behavior
of sections obtained by adding the Mordell-Weil generator multiple times.

In section \ref{ss:more}, we construct
the models $\sT_n$, $n=0,\cdots,6$
and examine their properties.
We first discover these models within the
context of a general construction of threefolds
of Mordell-Weil rank-one,
explained in appendix \ref{ap:2sect}.
We then verify the anomaly coefficients and
matter charges of these models using their
enhancement to $SU(2)$ models.
We conclude with verifying the matter charges
by explicitly checking the intersection numbers
of the Mordell-Weil generator with the
isolated rational curves in the manifold.
As expected, there are two classes of isolated rational
curves, each contributing charge 1 and 2
hypermultiplets to the six-dimensional spectrum
respectively.

\subsection{$\sT_0$} \label{ss:beq6}

In this section, we study in detail the 
Calabi-Yau threefold $\Xh_0$ that yields 
the six-dimensional $T=0$ supergravity
theory with gauge group $U(1)$ and $b=6$.
We derive the Weierstrass model and
work out the resolution that maps
the singular model to the smooth manifold.
We also identify the Mordell-Weil generator
of the fibration and identify the sections
obtained by adding the generator multiple times.
We end by computing the intersection number
between the sections obtained in this way
and the isolated rational curves of the manifold.

Recall that the  Weierstrass from of
a Calabi-Yau threefold fibered
over $\field{P}^2$ is given by
\be
X_0~:\quad y^2 = x^3 +F_{12} xz^4 + G_{18} z^6
\ee
as it is a fibration over $\field{P}^2$.
$F_{12}$ and $G_{18}$ are homogeneous polynomials
of degrees $12$ and $18$ with respect to the projective
coordinates of $\field{P}^2$.
As can be seen from section \ref{ss:teq0ab},
the Mordell-Weil generator of an elliptic fibration
with $b=6$ must be of the form
\be
\sh~:~[x,y,z]=[f_6,f_9,1]
\label{1sect}
\ee
as $6=2(0+3)$, {\it i.e.,} the Mordell-Weil
generator does not
intersect the zero section.
As before, the subscripts on the polynomials indicate
their degree in the $\field{P}^2$ coordinates.

The form of the section enables us to write
the Weierstrass form as
\be
X_0~:\quad (y-f_9)(y+f_9) = (x-f_6)(x^2+f_6x+f_{12})
\label{r1}
\ee
in local coordinates in the $z=1$ chart.
The discriminant is given by
\be
\Delta=
f_9^2 (27f_9^2-54f_6 f_{12})
+(f_{12}+2 f_6^2)^2
(4f_{12}-f_6^2) \,.
\ee
The singular points of $X_0$ are
located at
\be
y=0,\qquad
x=f_6,\qquad
f_9=0,\qquad
f_{12}+2f_6^2=0 \,.
\label{108p}
\ee
There are $12 \times 9 =108$ points satisfying
these conditions lying above the $108$ points
in the base satisfying the latter two equations.
Indeed, the Weierstrass equation can be
rewritten in the useful form
\be
y^2 - f_9^2
=(x+2f_6) (x-f_6)^2 + (f_{12}+2f_6^2)(x-f_6)\,,
\ee
which makes clear that there are conifold
singularities at these points.

We can resolve these 108 points by blowing up
a codimension-two locus to a single divisor
in the ambient space, thereby recovering the
smooth manifold $\Xh_0$.
This transition can be described by a birational map
\cite{MV2,Klemm:1996hh,Louis:1996mt}.
In order to explain this birational map,
it is useful to represent $X_0$ and $\Xh_0$
as hypersurfaces in projective varieties.
$X_0$ --- represented by the equation \eq{r1} ---
can be thought of as a singular degree $18$
hypersurface in $\field{P}[1,1,1,6,9]$
with projective coordinates $(a,b,c,x,y)$.
The $a,b$ and $c$ coordinates are the
projective coordinates of the base manifold.
We can resolve this manifold into a
smooth degree $12$ hypersurface $\Xh_0$
in $\field{P}[1,1,1,3,6]$.
We denote the projective coordinates
of this manifold by $(a,b,c,v,w)$.\footnote{In order to
keep track of all the sections properly,
one must actually use toric ambient varieties
with divisors representing the sections.
In this subsection, we proceed with
the current presentation for sake of convenience
and comment on the loci of sections when necessary.
We deal with these issues more carefully
in the next subsection.}
Then, the birational map from $X_0$
to $\Xh_0$ is given by
\be
v={y+f_9 \ov 2(x-f_6)},\qquad
w={1 \ov 2}(x+{f_6 \ov 2}) - v^2 \,.
\label{Xh0birat}
\ee
We may rewrite \eq{r1} as
\be
\Xh_0~:~
w^2 = v^4 -{3 \ov 2} f_6 v^2 - f_9 v
+ ({f_6^2 \ov 16}-{f_{12} \ov 4})
\label{Xh0}
\ee
in these coordinates.
This is a generic degree 12 hypersurface in
$\field{P}[1,1,1,3,6]$.

\begin{figure}[!t]
\centering\includegraphics[width=9cm]{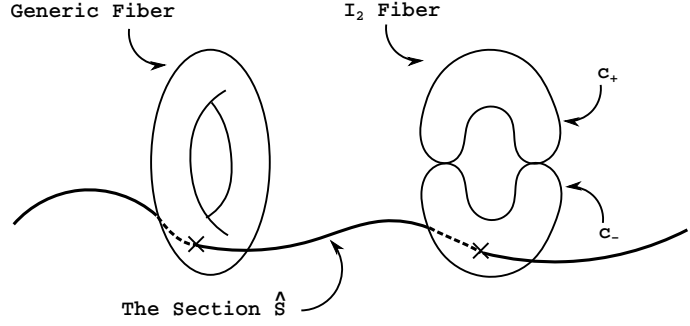}
\caption{\small The intersection between the section $\Sh$
and fiber components. $\Sh$ intersects a generic
fiber at a point. There are 108 loci in the base above
which the fiber becomes reducible --- in fact, an $I_2$ fiber.
The $I_2$ fiber consists of two
rational curves $c_+$ and $c_-$ intersecting at two points.
The section $\Sh$ intersects the $I_2$ fibers
at a point on the $c_-$ component.}
\label{f:s1}
\end{figure}

It is easy to see that the 108 singularities at \eq{108p}
are blown up into rational curves.
The fibers above the 108 loci given by
\be
f_9=0,\quad \fh_{12}\equiv f_{12}+2f_6^2=0
\ee
in the base, are resolved into $I_2$ fibers
\be
c^I_{\pm}~:~w=\pm \left( v^2-{3 \ov 4}f_6 \right) \,.
\label{I2}
\ee
The index $I=1,\cdots,108$ labels the loci
of the reducible fibers.
Each $I_2$ fiber consists of two rational
curves $c^I_+$ and $c^I_-$ intersecting at two points.
The isolated rational curve obtained by resolving the
singularity at each of the points is $c^I_-$,
while the zero section passes through the curve $c^I_+$.
The section \eq{1sect} intersects the 108 $c^I_-$ curves
at the point ``$v = \infty$" once.\footnote{More precisely,
the section \eq{sect} is given by $w/v^2=-1$. The zero
section is at $w/v^2=1$.}
Figure \ref{f:s1} depicts how $\Sh$ intersects the
fibral curves $c^I_-$.
Hence we have accounted for all the shrinking rational curves
--- they are given by $c^I_-$ and
\be
c^I_- \cdot \Sh = c^I_- \cdot \sigma(\sh) =1 \,.
\ee
Therefore the six-dimensional theory has 108 hypermultiplets
with unit charge under the $U(1)$ vector field dual to $S$.
This correctly reproduces data of the $b=6$ theory.

Now let us examine the sections generated by the section $\sh$
in this elliptically fibered manifold.
We denote the homology class of the section $m \sh$ by $\Sh_m$.
Through explicit calculation, we write down the following few sections
in mutually relatively prime
fiber coordinates $(x,y,z)$ of the Weierstrass representation:
\begin{align}
\begin{split}
-\sh~:~[x,y,z]&=[f_6,-f_9,1] \\
\sh~:~[x,y,z]&=[f_6,f_9,1] \\
2\sh~:~[x,y,z]&=\left[ \fh_{12}^2-8f_6 f_9^2,
-\fh_{12}^3+12f_6 \fh_{12} f_9-8 f_9^4,2f_9 \right] \\
3\sh~:~[x,y,z]&=\left[F_{54},
F_{81}, \fh_{12}^2-12f_6f_9^2  \right] \\
&\vdots
\end{split}
\label{sections}
\end{align}
Recall that we have defined $\fh_{12}=f_{12}+2f_6^2$.
$F_{54}$ and $F_{81}$ are order $54$ and $81$
polynomials that we have not written out explicitly.
It is satisfying to check that the orders of these polynomials
are indeed given by
\be
\Sh_{m}~:~[x,y,z]=\left[F_{6m^2}, F_{9m^2}, F_{3m^2-3}  \right] 
\ee
as predicted by equation \eq{multorder}.

Since $\Sh \cdot c=\sigma(\sh) \cdot c=1$
for the 108 fibral rational
curves in the resolved manifold,
it follows that
\be
\Sh_m \cdot c = \sigma(m\sh) \cdot c =
m\sigma(\sh) \cdot c =m \,.
\label{multint}
\ee
This implies that a section can have
arbitrary intersection numbers with rational curves.
Let us verify these intersection numbers for
$m=-1$ and $2$ for the rest of this subsection.
The example $m=2$ turns out to be useful in
analyzing models with $b>6$.

In order to verify the intersection numbers between
sections and fibral curves in the resolved manifold $\Xh_0$,
it is convenient to view it as a resolution of a
determinantal variety:
\be
M \bpm V \\ T \epm
=
\bpm
x-f_6 & y+f_9 \\
y-f_9 & (x^2+f_6 x+ f_{12}) \\
\epm
\bpm V \\ T \epm
=
\bpm 0 \\ 0 \epm
\label{X0prime}
\ee
Here, $T$ and $V$ are projective coordinates of a $\field{P}^1$.
Away from the $108$ singular loci, \eq{X0prime}
is solved by
\begin{align}
(y-f_9)(y+f_9) &= (x-f_6)(x^2+f_6x+f_{12}) \\
V:T=-(y+f_9):(x-f_6) &= -(x^2+f_6 x+f_{12}):(y-f_9)
\end{align}
As the matrix $M$ has rank-one at non-singular points of $X_0$,
a unique point on $\field{P}^1$ is assigned to every non-singular
point of $X_0$. Meanwhile, at the $108$ singular points \eq{108p},
the matrix $M$ becomes rank zero --- the singular point is replaced
by the full $\field{P}^1$ parametrized by $V/T$.
In fact, the coordinate $v$ used in the birational map \eq{Xh0birat}
is a coordinate on this $\field{P}^1$:
\be
v=-{V\ov 2T} \,.
\ee
The coordinate $w$ \eq{Xh0birat} is a linear combination
of $v^2$ and $x$.
For the purpose of computing intersection numbers of
fibral curves and sections, it is more convenient
to use the local coordinates $v$ and $x$ rather than
$v$ and $w$.

The resolved fibral curves
$c^I_-$ sitting above the loci $f_9=\fh_{12}=0$
in the base can be written as
\be
c^I_{-} ~: \quad x=f_6 \,,
\label{cI-res}
\ee
with unrestricted $V/T$.
The other component $c^I_+$ of the $I_2$ fiber
is given by
\be
c^I_{+} ~: \quad x=4v^2-2f_6 \,.
\label{cI+res}
\ee

\subsubsection{$m=-1$}
\begin{figure}[!t]
\centering\includegraphics[width=9cm]{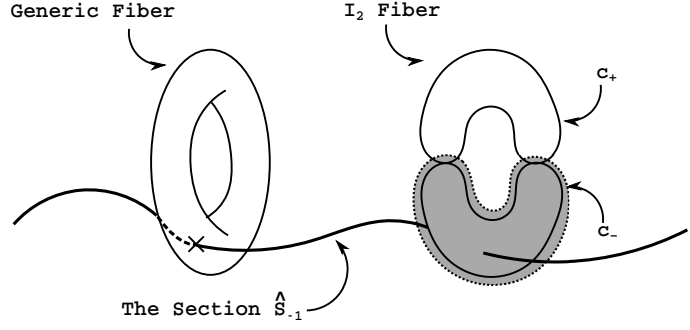}
\caption{\small The section $\Sh_{-1}$.
$\Sh_{-1}$ intersects a generic
fiber at a point while is resolved into the $c_-$
component at the $I_2$ loci.}
\label{f:sm1}
\end{figure}

Let us examine the section
\be
-\sh~:~[x,y,z]=[f_6,-f_9,1] \,.
\ee
Plugging in the locus of this section
to equation \eq{X0prime} we find that
$\field{P}^1$ coordinates are given by
\be
\bpm
0 & 0 \\
-2f_9 & \fh_{12} \\
\epm
\bpm V \\ T \epm
=
\bpm 0 \\ 0 \epm \,.
\ee
Therefore at the loci $f_9=\fh_{12}=0$
in the base, the locus of the section on the fiber
becomes
\be
x=f_6 \,,
\ee
with unrestricted $V/T$, {\it i.e.,}
it is blown up into $c^I_-$.
This behavior is depicted in figure \ref{f:sm1}.

A quick way to compute the intersection
numbers between the homology class
$\Sh_{-1}$ of the section $-\sh$
and the fibral curves $c^I_-$ is the following.
Since $c^I_+$ and $c^I_-$ intersect
at two points,
\be
\Sh_{-1} \cdot c^I_+ =2 \,,
\ee
as $\Sh_{-1}$ is resolved into $c^I_-$ at
locus $I$.
Meanwhile, $\Sh_{-1}$, being a homology class of
the section satisfies
\be
\Sh_{-1} \cdot f=1
\ee
for the fiber class $f$. Using the fact that
\be
f=c^I_- + c^I_+
\ee
for each $I=1,\cdots,108$, we find that
\be
\Sh_{-1} \cdot c^I_- =-1 \,,
\ee
which indeed confirms \eq{multint}.

\subsubsection{$m=2$} \label{sss:meq2}

Let us consider the section given by
\be
2\sh~:~[x,y,z]=\left[ \fh_{12}^2-8f_6 f_9^2,
-\fh_{12}^3+12f_6 \fh_{12} f_9-8 f_9^4,2f_9 \right]
\label{sh2}
\ee
in projective coordinates.
Again, plugging in the locus of this section
\be
x=\left({\fh_{12} \ov 2 f_9}\right)^2-2f_6,\quad
y=-\left({\fh_{12} \ov 2 f_9}\right)^3 +3f_6 \left({\fh_{12} \ov 2 f_9}\right)^2 -f_9
\ee
--- in the chart $z=1$ ---
to equation \eq{X0prime} we find that
the matrix $M$ is given by
\be
M= \bpm
\left({\fh_{12} \ov 2 f_9}\right)^2-3f_6 ~&~ -\left({\fh_{12} \ov 2 f_9}\right) \left\{\left({\fh_{12} \ov 2 f_9}\right)^2-3f_6 \right\} \\
-\left({\fh_{12} \ov 2 f_9}\right)^3+3f_6\left({\fh_{12} \ov 2 f_9}\right)-2f_9~&~
-\left({\fh_{12} \ov 2 f_9}\right)\left\{-\left({\fh_{12} \ov 2 f_9}\right)^3+3f_6\left({\fh_{12} \ov 2 f_9}\right)-2f_9 \right\}\\
\epm \,.
\ee
Therefore the projective coordinates of the
$\field{P}^1$ are given by
\be
V:T = \fh_{12} : 2f_9
\ee
for points of the section above
non-degenerate loci on the base.

\begin{figure}[!t]
\centering\includegraphics[width=9cm]{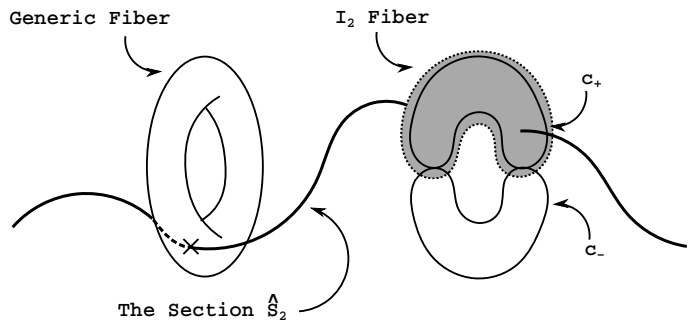}
\caption{\small The section $\Sh_{2}$.
$\Sh_{2}$ intersects a generic
fiber at a point while is resolved into the $c_+$
component at the $I_2$ loci.}
\label{f:s2}
\end{figure}

At the loci where the fiber becomes degenerate
($f_9=\fh_{12}=0$), the section
is not well-defined as all the projective
coordinates in \eq{sh2} become zero.
Therefore one must resolve the section at these loci
to treat them correctly in the blown-up manifold.
This can be done globally by introducing the $\field{P}^1$
coordinates $(P,Q)$ along the section such that
\be
f_9P-\fh_{12}Q=0
\quad \Leftrightarrow \quad
p= \fh_{12}/f_9 \,.
\ee
We note that this resolution of the section
does not introduce any new divisor in the ambient
space. Rather, it attaches curves to the section
within the resolved ambient space to produce a
``closure" of the section.

As a result, the section $2\sh$ is resolved into
\be
v=-p, \quad x=4p^2-2f_6
\ee
at the $I_2$ loci.
This behavior is depicted in figure \ref{f:s2}.
For each locus $I=1,\cdots,108$,
this is precisely the curve $c^I_+$ by \eq{cI+res}.
Since $c^I_+$ intersects the curve $c^I_-$
at exactly two points, and since the
section $\Sh_2$ is resolved into
$c^I_+$ at each $I$,
\be
\Sh_2 \cdot c^I_-=2
\ee
for the fibral curves.
This result is consistent with \eq{multint}.

\subsection{$\sT_n$, $0 \leq n \leq 6$} \label{ss:more}

In this section, we study the 
Calabi-Yau threefolds $\Xh_n$ that yield
the theories $\sT_n$ with $n =0,\cdots,6$.
These are six-dimensional $T=0$ supergravity
theories with gauge group $U(1)$ and $b=2(n+3)$.
Recall that the matter content of $\sT_n$
is given by $4(n+3)(9-n)$ hypermultiplets with charge $1$
and $n(n+3)$ hypermultiplets with charge $2$.
We first derive the Weierstrass models $X_n$
using a general construction explained in
detail in appendix \ref{ap:2sect}.
We first check that these manifolds indeed
yield $\sT_n$ indirectly by field theory arguments.
We then explicitly verify that the F-theory
compactification upon $\Xh_n$ results in
$\sT_n$ by studying the Mordell-Weil generator
and its intersection numbers with fibral rational curves.

In appendix \ref{ap:2sect} it is shown that a
Weierstrass model of an elliptic fibration over field $K$
with Mordell-Weil group of rank one is of the form
\be
y^2=x^3 + (c_1 c_3 -b^2 c_0 - {c_2^2 \ov 3} ) x z^4
+\left( c_0c_3^2  -{1 \ov 3} c_1c_2c_3 +{2 \ov 27} c_2^3
-{2 \ov 3} b^2 c_0c_2 +{b^2 c_1^2 \ov 4} \right) z^6
\label{wsre}
\ee
with the Mordell-Weil generator
\be
[x,y,z]=[c_3^2-{2 \ov 3}b^2 c_2, -c_3^3+b^2 c_2 c_3 -{1 \ov 2} b^4 c_1,b] \,.
\label{rank1sectre}
\ee
Here $c_i$ and $b$ are elements of $K$.

There is a straightforward way of utilizing the
equation \eq{wsre} to obtain a class of
Weierstrass models of Calabi-Yau threefolds
fibered over $\field{P}^2$ with Mordell-Weil rank-one.
It is to set
\begin{align}
c_3 = f_{3+n}, \quad
c_2 = 3f_6, \quad
c_1 = 2f_{9-n}, \quad
c_0 = f_{12-2n}, \quad
b = b_{n}
\end{align}
where $f$ and $b$ are polynomials
of the base $\field{P}^2$ coordinates
whose subscripts denote their degree.
The proportionality constants are designated
for aesthetic reasons.
Under these assignments the Weierstrass form
\eq{wsre} becomes
\begin{align}
\begin{split}
X_n~:\quad
y^2&=x^3+(2f_{3+n}f_{9-n}-3f_6^2-b_n^2 f_{12-2n})xz^4 \\
&+(2f_6^3-2f_{3+n}f_6f_{9-n}+f_{3+n}^2 f_{12-2n}-2b_n^2f_6 f_{12-2n} +{b_n^2 f_{9-n}^2})z^6 \,.
\label{rank1ansatz}
\end{split}
\end{align}
The Mordell-Weil generator of the fibration is
given by
\be
\sh~: ~[x,y,z]=[f_{3+n}^2-2b_n^2 f_6, -f_{3+n}^3+3b_n^2 f_6 f_{3+n} -b_n^4f_{9-n},b_n]
\label{rank1ansatzsect}
\ee
in mutually relatively prime projective coordinates.\footnote{
We explicitly verify that this section is a Mordell-Weil generator
shortly, using charge minimality conditions discussed at
the end of section \ref{ss:teq0ab}.}
Upon compactifying F-theory on these manifolds, one
obtains $T=0$ theories with gauge group $U(1)$.
From the form of the Mordell-Weil generator,
it follows that the anomaly coefficient of the $U(1)$ theory is
\be
b=2(n+3)
\label{bval}
\ee
as explained in section \ref{ss:teq0ab}.

We claim that for each $n$, the low-energy theory
obtained by compactifying F-theory on $X_n$
is $\sT_n$.
For the rest of this section, we verify this claim
by using field theory arguments (section \ref{sss:fieldtheory})
and by direct computation of intersection numbers in the
resolution of $X_n$ (section \ref{sss:directcomputation}).

We note that the ansatz \eq{rank1ansatz}
is valid only for $0 \leq n \leq 6$
by the explicit form of the Weierstrass model
--- since there is polynomial of degree $f_{12-2n}$, $n$
cannot exceed $6$.
There is an obvious extension of this ansatz allowing $f_{12-2n}$
to vanish identically if $n>6$, but as we observe in
appendix \ref{ap:C} the Weierstrass model
acquires an additional unintended $SU(2)$ gauge factor
in the extended ansatz.

\subsubsection{Field Theory} \label{sss:fieldtheory}

A quick way of deriving the low-energy theory
of $X_n$ is by Higgsing.
As commented at the end of appendix \ref{ap:2sect},
upon tuning $b_n \ra 0$, $X_n$ becomes
\begin{align}
\begin{split}
y^2=x^3+(2f_{3+n}f_{9-n}-3f_6^2)x z^4
+(2f_6^3-2f_{3+n}f_6f_{9-n}+f_{3+n}^2 f_{12-2n}) z^6\,.
\label{enhanced}
\end{split}
\end{align}
The discriminant locus of this model is given by
\be
f_{3+n}^2 \left\{
36f_6^2 \left( 3f_6 f_{12-2n}\!-\!f_{9-n}^2 \right)
\!+\!f_{3+n} \left( 32f_{9-n}^3 \!-\!108 f_6 f_{9-n} f_{12-n}
\!+\!27 f_{3+n}f_{12-2n}^2 \right) \right\}
\label{disc}
\ee
This theory is an $SU(2)$ theory that has an enhanced
gauge group over $f_{3+n}=0$.
We have un-Higgsed the $U(1)$ theory to an $SU(2)$
theory by tuning the hypermultiplets by $b_n \ra 0$.

Let us examine the properties of the $SU(2)$ model.
The anomaly coefficient of the $SU(2)$ group is $(n+3)$,
and the $SU(2)$ locus in the base has genus $(n+2)(n+1)/2$.
Therefore there are $(n+2)(n+1)/2$ adjoint hypermultiplets
in the low-energy spectrum of the $SU(2)$ theory.
From the discriminant locus \eq{disc}
one finds that there are $2(n+3)(9-n)$ fundamental
hypermultiplets localized at the loci
\be
f_{3+n}=0, \quad 3f_6 f_{12-2n}-f_{9-n}^2=0
\ee
in the base, where the $I_2$ fiber enhances to
an $I_3$ fiber.
We note that there are not any additional matter
localized at
\be
f_{3+n}=0, \quad f_6=0
\ee
as the fiber reduces to a type $III$ fiber at these loci.
The mixed/gauge anomaly equations are satisfied for
this theory:
\be
\GG=SU(2)~:\quad
2(n+3)(9-n) \times {\tiny\yng(1)} +
{(n+2)(n+1) \ov 2} \times ({\rm Adj}),
\qquad b_{SU(2)} = (n+3) \,.
\ee

The $U(1)$ theory given by the manifold $X_n$
\eq{rank1ansatz} with Mordell-Weil rank-one
can be obtained by Higgsing the $SU(2)$ theory
in a particular way.
There are $J\equiv (n+2)(n+1)/2$ adjoint fields
in the $SU(2)$ theory.
Turning on these fields ($\Phi_1, \Phi_2, \cdots, \Phi_J$)
to a generic value will completely break the gauge
symmetry while turning them on such that
\be
\Phi_1 = c_1 \sigma_3,~
\Phi_2 = c_2 \sigma_3,~
\cdots,~
\Phi_J = c_J \sigma_3 \,,
\ee
breaks the theory to a $U(1)$ theory.
The $J=(n+2)(n+1)/2$ parameters $c_1,\cdots, c_J$ are
encoded in the $(n+2)(n+1)/2$ coefficients of
the polynomial $b_n$.

The $U(1)$ theory obtained by Higgsing the adjoint
hypermultiplets in this way has $4(n+3)(9-n)$
charge $1$ hypermultiplets coming from the fundamental
fields and $n(n+3)$ charge $2$ hypermultiplets
coming from the adjoint fields.
Its anomaly coefficient is twice of that of the
$SU(2)$ theory, as the normalized coroot matrix
of $SU(2)$ is $(2)$, {\it i.e.,}
\be
b = 2b_{SU(2)} = 2(n+3) \,,
\ee
which is consistent with \eq{bval}.
This is precisely the theory $\sT_n$, as claimed.

\subsubsection{Direct Computation} \label{sss:directcomputation}

\begin{table}[!t]
\centering
\begin{tabular}{c|cccccc}
\hline
$\field{C}^*$ &  $a$ &$b$ &$c$ & $x$ &$y$ &$z$\\
\hline
1 &  $1$ &$1$ &$1$ & $0$ &$0$ &$-3$ \\
2 &  $0$ &$0$ &$0$ & $2$ &$3$ &$1$ \\
\hline
\end{tabular}
\caption{The toric data of the ambient space of
manifold $X_n$. $a,b$ and $c$ are the $\field{P}^2$
coordinates, while $x, y$ and $z$ are the fiber coordinates.}
\label{t:toric1}
\end{table}  

Let us verify that $X_n$ yields $\sT_n$ directly from the geometry.
We proceed by first resolving $X_n$ to a smooth threefold $\Xh_n$.
We then work out the resolution of the section \eq{rank1ansatzsect}
under this map and compute its intersection numbers with
the fibral rational curves, thereby confirming the
charges of the hypermultiplets of the theory.

Let us begin by noting that the Weierstrass model
\eq{rank1ansatz},
\begin{align}
\begin{split}
X_n~:\quad
y^2&=x^3+(2f_{3+n}f_{9-n}-3f_6^2-b_n^2 f_{12-2n})xz^4 \\
&+(2f_6^3-2f_{3+n}f_6f_{9-n}+f_{3+n}^2 f_{12-2n}-2b_n^2f_6 f_{12-2n} +{b_n^2 f_{9-n}^2})z^6 \,,
\nonumber
\end{split}
\end{align}
is a hypersurface of a toric variety
whose coordinates and $\field{C}^{*}$ actions
are summarized in table \ref{t:toric1} (describing a
$\mathbb{P}^{(1,2,3)}$-bundle over $\mathbb{P}^2$).
The smooth threefold $\Xh_n$ birationally equivalent to $X_n$
is given by a  hypersurface of a
toric variety $T_n$ whose data can be summarized 
by\footnote{We thank Christoph Mayrhofer
for assistance in identifying these resolved manifolds.}
table \ref{t:toric2}
(describing a bundle over $\mathbb{P}^2$ whose fiber is
$\operatorname{Bl}_{[0,1,0]}\mathbb{P}^{(1,1,2)}$ --- see
appendix \ref{ap:2sect}).

The birational map between $X_n$ and $\Xh_n$
is given by a slight modification of equation \eq{e7toe8}.
Let us examine this map in detail.
A useful way of describing the map is to
identify $\field{C}^{*}$ actions of the two toric ambient
spaces. We identify
$\field{C}^{*}_1$/$\field{C}^{*}_2$ of table \ref{t:toric1} with
$\field{C}^{*}_{1'}$/$\field{C}^{*}_{2'}$ of table \ref{t:toric2}
respectively.
Then the birational map \eq{e7toe8}
can be described in terms of $\field{C}^{*}_{3'}$
invariant coordinates.
Defining the $\field{C}^{*}_{3'}$ invariant coordinates
of $\Xh_n$ as
\be
u := U, \quad v :=V/T, \quad w :=W/T \,,
\ee
the birational map is given by a reparametrized
version of \eq{e7toe8}\footnote{The reparametrization is
obtained by replacing $w$ of \eq{e7toe8}
by $w-bv^2$.}:
\begin{align}
\begin{split}
x &= f_{3+n} uv + f_{6} u^2 + b_n w\\
y &= b_n f_{3+n} uv^2 + 3b_n f_6 u^2 v
+ { b_n f_{9-n} } u^3 +f_{3+n} uw+b_n^2 vw\\
z &= u
\end{split}
\label{nbirat}
\end{align}
The inverse map is given by
\begin{align}
\begin{split}
v &= {b_n (y-f_{9-n}b_nz^3)-f_{3+n} (x-f_6 z^2)z
\ov b_n^2(x+2f_6 z^2)-f_{3+n}^2z^2}\\
w &=
{-f_{3+n} (y-f_{9-n}b_nz^3)z +b_{n} (x+2f_6z^2)(x-f_6 z^2)
\ov b_n^2(x+2f_6 z^2)-f_{3+n}^2z^2}\\
u &= z
\end{split}
\label{invbirat}
\end{align}
Under this birational map, $X_n$ is mapped to
\be
\Xh_n ~: \quad TW^2 -b_n WV^2 = U \left(
f_{3+n} V^3 + 3f_6 TUV^2
+2f_{9-n} T^2 U^2 V + f_{12-2n} T^3 U^3
\right)
\,,
\label{Xhn}
\ee
which is a generic hypersurface in the toric variety
$T_n$ when $n \leq 6$.

\begin{table}[!t]
\centering
\begin{tabular}{c|ccccccc}
\hline
$\field{C}^*$ &  $a$ &$b$ &$c$ & $T$ &$U$ &$V$ &$W$\\
\hline
$1'$ &  $1$ &$1$ &$1$ & $n$ &$-3$ &$0$ &$0$\\
$2'$ &  $0$ &$0$ &$0$ & $0$ &$1$ &$1$ &$2$\\
$3'$ &  $0$ &$0$ &$0$ & $1$ &$0$ &$1$ &$1$\\
\hline
\end{tabular}
\caption{The toric data of the ambient space $T_n$ of
manifold $\Xh_n$. $a,b$ and $c$ are the $\field{P}^2$
coordinates. $T, U, V$ and $W$ are the fiber coordinates.}
\label{t:toric2}
\end{table}

All the fibral rational curves of $\Xh_n$
are isolated, as there are no fibral divisors.
These rational curves are components of $I_2$
fibers. The fiber degenerates to $I_2$ fibers at
codimension-two loci above the base in $\Xh_n$.
There are two different types of $I_2$ loci ---
charge-two loci and charge-one loci,
where isolated rational curves that contribute hypermultiplets
of charge-two and one to the six-dimensional
spectrum are localized, respectively.
There are $n(n+3)$ charge-two loci
and $4(n+3)(9-n)$ charge-one loci,
as we expect from the preceding discussions.

Let us examine the charge-two loci.
From the defining equation \eq{Xhn} it is easy
to see that the fiber degenerates at the
$n(n+3)$ codimension-two loci in the base
\be
b_n =0, \quad  f_{3+n}=0 \,.
\label{ctl}
\ee
At these points, \eq{Xhn} becomes
\be
TW^2= 3f_6  T U^2 V^2
+2f_{9-n}T^2 U^3 V + f_{12-2n} T^3 U^4 \,.
\ee
The two rational curves that consist the $I_2$
fiber are
\begin{align}
\begin{split}
\chi^\iota_+~&: \quad T =0 \\
\chi^\iota_-~&: \quad W^2 =3 f_6  U^2 V^2 +2 f_{9-n}T U^3 V + f_{12-2n} T^2 U^4
\end{split}
\label{chi}
\end{align}
at each point indexed by $\iota=1, \cdots, n(n+3)$.
We call these loci ``charge-two loci,'' as the fibral
rational curves $\chi_-$ sitting above these points contribute
hypermultiplets of charge $2$ under the $U(1)$.
We can see that the degenerate fiber is indeed an $I_2$
fiber as the two curves $\chi^\iota_+$ and $\chi^\iota_-$
meet at the two points
\be
{W \ov UV} = \pm \sqrt{3f_6} \,.
\ee
To verify that $\chi_-$ are the fibral curves blown down
by the map \eq{nbirat}, we can plug in $b_n =0,  f_{3+n}=0$
to this formula to see that
\be
x=f_6 u^2,\quad y=0, \quad z=u
\ee
when $T \neq 0$.
These are exactly the projective coordinates
of the singular point of $X_n$ at the charge-two
locus $b_n =0,  f_{3+n}=0$, as the Weierstrass
model reduces to
\begin{align}
y^2&=(x+2f_6z^2)(x-f_6z^2)^2
\end{align}
at these points.

Let us verify that the intersection numbers
between the Mordell-Weil generator
\be
\sh~: ~[x,y,z]=[f_{3+n}^2-2b_n^2 f_6,-f_{3+n}^3+3b_n^2 f_6 f_{3+n} -b_n^4f_{9-n},b_n] \,,
\label{sect}
\ee
and the isolated rational curves $\chi^\iota_-$
at the charge-two loci are indeed given by
\be
\sigma(\sh) \cdot \chi^\iota_- = \Sh \cdot \chi^\iota_- =2 \,.
\ee
Plugging in the Weierstrass coordinates
of the section \eq{sect} into the map \eq{invbirat},
we find that the section maps to the point
\be
T=0,\quad {W \ov UV} =-{f_{3+n} \ov b_n}
\label{intpoint}
\ee
above generic points in the base.
The section, however,
is not well-defined at the charge-two loci \eq{ctl}.
As in section \ref{sss:meq2}, we resolve these points on the
section, {\it i.e.,} we let
\be
b_n P - f_{3+n} Q=0 \,,
\ee
where $(P,Q)$ parametrizes a $\field{P}^1$.
By this resolution, the section $\sh$ is resolved
to the curve $\chi^\iota_+$
at the charge-two loci.
Therefore, the section $\sh$ intersects
the curves $\chi^\iota_-$ at two points.

The $I_2$ fibers other than the charge-two
fibers can be found in the following way.
Using the $\field{C}^*_{3'}$ invariant coordinates,
the equation for $\Xh_n$ can be written in the form
\be
(w-{b_n \ov 2}v^2)^2 =
{1 \ov 4} b_n^2 v^4
+f_{3+n} v^3 + 3f_6 v^2
+2f_{9-n} v + f_{12-2n} \,,
\label{c1}
\ee
where we have  set $u =1$.
The $I_2$ loci other than the charge-two
loci are the codimension-two points in the base where
the right-hand-side of this equation
factors into
\be
({1 \ov 2} b_n v^2
+{f_{3+n} \ov b_n} v
+ {3b_n^2 f_6-f_{3+n}^2 \ov b_n^3})^2 \,.
\label{c1p}
\ee
These are the charge-one loci.
By equating \eq{c1} and \eq{c1p}
we find that the charge-one loci are given by
the points that satisfy
\begin{align}
\begin{split}
f_{3+n}^3-3 f_6 f_{3+n} b_n^2+b_n^4 f_{9-n} &=0 \\
f_{3+n}^4-6 f_6 f_{3+n}^2 b_n^2+9 f_6^2 b_n^4 -f_{12-2n}b_n^6 &=0
\end{split}
\label{c1loci}
\end{align}
that are not charge-two loci.
For a generic $\Xh_n$,
neither $b_n$ nor $f_{3+n}$
vanishes at a charge-one locus.

Near the charge-one loci, the resolution
\eq{e7toe8} and the section \eq{sect} exhibit the
same behavior as in $\sT_0$, which we have
extensively studied in section \ref{ss:beq6}.
To verify this behavior it proves useful to define
\be
p_3 \equiv {f_{3+n}/b_n} \,.
\ee
The Weierstrass model $X_n$
can be written as
\be
y^2-g_9^2=(x-g_6)(x^2+g_6x+g_{12})
\ee
for
\begin{align}
\begin{split}
g_6=p_3^2-2 f_6, \quad
g_9 = -p_3^3+3f_6p_3 -b_n f_{9-n}, \\
g_{12}=p_3^4-4f_6 p_3^2+2b_n f_{9-n} p_3+f_6^2 -b_n^2f_{12-2n}\,,
\end{split}
\end{align}
where we have set $z=1$.
The Mordell-Weil generator \eq{sect}
is given by
\be
\sh~: ~[x,y,z]=[g_6,g_9,1] \,.
\ee
Also, the birational map \eq{invbirat} can be re-written
in the form
\begin{align}
\begin{split}
b_n v &= {y+g_9 \ov x-g_6} -p_3 \\
b_n w &= -p_3 b_n v +(x+{g_6 \ov 2} -{p_3 ^2 \ov 2})
\end{split}
\end{align}
The charge-one loci \eq{c1loci}
are the loci at which
\begin{align}
\begin{split}
g_9 &= -p_3^3+3f_6p_3 -b_n f_{9-n} = 0 \\
\hat{g}_{12} &\equiv g_{12} + 2 g_6^2  =
3p_3^4-12f_6p_3^2 + 2b_n f_{9-n} p_3 +9f_6^2-b_n^2f_{12-2n}=0
\end{split}
\end{align}
From the fact that
$b_n$, $f_{n+3}$ and $p_3$ are all well-defined
and non-zero at the charge-one loci,
it is clear that the analysis of
section \ref{ss:beq6} can be readily applied
to understanding these points.

The singular fibers of $X_n$ located
above these points are resolved into $I_2$ fibers
that consist of two rational curves:
\be
c^I_{\pm}~:~w-{b_n \ov 2}v^2
=\pm \left( {1 \ov 2}b_n v^2 +p_3v +{p_3^2-3g_6 \ov 2 b_n} \right) \,.
\label{c1I2}
\ee
We have used $I$ to index the
charge-one loci.
Using the full set of projective coordinates,
these curves can be written as
\be
c^I_{\pm}~:~TW-{b_n \ov 2}V^2
=\pm \left( {1 \ov 2}b_n V^2 +p_3TUV +{p_3^2-3g_6 \ov 2 b_n} T^2U^2 \right) \,.
\ee
The zero-section,
\be
U=0, \quad {TW \ov V^2} =b_n \,,
\ee
intersects the curve $c^I_+$ at a single point,
while the Mordell-Weil generator \eq{intpoint},
\be
T=0,\quad {W \ov UV} = -p_3 \,,
\ee
intersects $c^I_-$ also at a single point.
Therefore the isolated rational curve at the charge-one
locus $I$ is $c^I_-$ and its intersection number
with the Mordell-Weil generator is indeed given by
\be
\sigma(\sh) \cdot c^I_- = \Sh \cdot c^I_- =1 \,.
\ee
It is clear that $\sh$ is a Mordell-Weil generator, as there
exist curves of unit intersection number with $\Sh$.

Let us now show that there are $4(n+3)(9-n)$
charge-one loci.
To count the number of charge-one loci,
one must count the number of points that
satisfy \eq{c1loci}, but at which $b_n\neq 0$
and $f_{n+3} \neq 0$.
To show that there are $4(n+3)(9-n)$
such points, it is enough to show this
in the case when
\be
b_n = \epsilon B_n
\ee
for $\epsilon$ that is in a small
neighborhood of $0$.
To do so, let us first rewrite
\eq{c1loci} as
\begin{align}
\begin{split}
p_3^3 -3f_6 p_3 + b_n f_{9-n} &=0 \\
p_3^4 -6f_6 p_3^2 +9f_6^2 -b_n^2 f_{12-2n} &=0 \\
\end{split}
\end{align}
and view these equations as
polynomial equations with respect to $p_3$.
These two equations can have
a unique common root
\be
p_3={b_n f_{9-n} f_{12-2n} \ov 3f_6 f_{12-2n} -f_{9-n}^2}
\label{constr1}
\ee
only when
\be
(f_{9-n}^2-3f_6 f_{12-2n})^2=b_n^2f_{12-2n}^3 \,.
\label{constr2}
\ee
When $|\epsilon| <<1$,
\be
f_{n+3} = {f_{9-n}(3f_6 f_{12-2n} -f_{9-n}^2) \ov f_{12-2n}^2}=\OO (\epsilon) \,.
\ee
At small enough $\epsilon$,
for each point satisfying the two equations
\eq{constr1} and \eq{constr2},
there exists a nearby point satisfying
\be
f_{n+3} = 0 \,,
\ee
along with \eq{constr2}.
Therefore when $\epsilon$ lies in a small
enough neighborhood of $0$, there are
$(n+3)(36-4n)$ charge-one loci.
We note when the theory is enhanced to an
$SU(2)$ theory by taking
$\epsilon \ra 0$, the charge-one loci
merge in pairs to the codimension-two points
defined by
\be
f_{n+3} = 0, \quad f_{9-n}^2-3f_6 f_{12-2n}=0\,.
\ee
These are precisely the points above which the
fundamental hypermultiplets of the enhanced $SU(2)$
sit.

\begin{figure}[!t]
\centering\includegraphics[width=12cm]{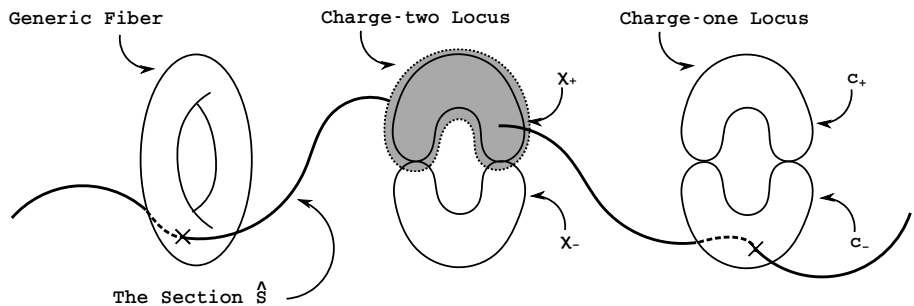}
\caption{\small The intersection between the section $\Sh$
and fiber components in $\Xh_n$.
$\Sh$ intersects a generic
fiber at a point. There are $n(n+3)$ charge-two loci 
and $4(n+3)(9-n)$ charge-one loci in the base above
which the fiber degenerates into an $I_2$ fiber.
The fibral curves $\chi_-^\iota$ localized above the charge-two
loci intersect the section $\Sh$ twice,
while the fibral curves $c_-^I$ at the charge-one loci
intersect $\Sh$ once.}
\label{f:tn}
\end{figure}

We have shown that there are
two types of isolated fibral rational curves in $\Xh_n$ ---
those localized above charge-two loci and
and those localized above charge-one loci in the base.
The $n(n+3)$ charge-two rational curves
intersect the Mordell-Weil generator $\Sh$
twice while the $4(n+3)(9-n)$
charge-one rational curves intersect
$\Sh$ once.
We have summarized these facts in figure \ref{f:tn}.

\section{Questions and Future Directions}  \label{s:directions}

There are a host of questions regarding
the Mordell-Weil group of elliptically fibered Calabi-Yau
threefolds that we have not pursued in this note.
We conclude by listing some interesting questions
--- in what we believe is to be the order of increasing difficultly ---
that could hopefully be addressed in the not-so-distant future.
\vspace{0.1in}

\noindent
{\bf Models with $n=7, 8$} \\
Out of the nine $T=0$, $U(1)$ theories
with hypermultiplets of charges $\leq 2$
allowed by anomaly equations,
we have only constructed F-theory
models for seven.
Although we have not been able to construct
the two theories --- which we have denoted
$\sT_7$ and $\sT_8$ in section \ref{s:anom} ---
we expect these theories to be embeddable
in F-theory.

This expectation comes from the fact that
the $U(1)$ theories $\sT_n$
can be obtained by Higgsing the adjoint
hypermultiplets of an $SU(2)$ theory with
$2(n+3)(9-n)$ fundamental hypermultiplets
and ${(n+2)(n+1)/2}$ adjoint hypermultiplets.
Let us denote the $SU(2)$ theories by $\sT_n'$.
The non-abelian theory $\sT_n'$ has anomaly
coefficient $b=10$($b=11$) and $88$($97$) neutral
hypermultiplets for $n=7$($n=8$) respectively. 
We are not aware of any obstruction in embedding
these un-Higgsed $SU(2)$ theories into F-theory
as they satisfy all the anomaly constraints and
also the Kodaira constraint, {\it i.e.,} $b<18$
for both these models.
It is on these grounds that
we expect $\sT_{7,8}'$ --- and
$\sT_{7,8}$, which can be obtained from
$\sT_{7,8}'$ by Higgsing ---
to be embeddable in F-theory.

We have not, however, been able explicitly
construct the threefolds that yield $\sT_{7,8}'$,
let alone $\sT_{7,8}$.
The difficulty in constructing these theories
originates from the fact that the ring of
polynomials with two variables is
not a Euclidean ring.
This implies that the ans\"atz
\eq{rank1ansatz}, \eq{enhanced} we have
used to construct Weierstrass models for
$\sT_{n}$, $\sT_{n}'$ do not necessarily
generalize to all possible $n$.

It would be interesting to explicitly
construct $\sT_{7,8}$/$\sT_{7,8}'$
or prove that they cannot be engineered
as F-theory models. It would be intriguing
if $\sT_{7}'$ or $\sT_8'$
defies expectations and is shown to be
un-embeddable in F-theory.
This would imply that the Kodaira constraint
is not a sharp enough criterion for discerning
whether a non-abelian theory can be embedded
in F-theory or not.
\vspace{0.1in}

\noindent
{\bf Models with General Charges} \\
A natural question to ask in light of the results
of this note is whether there exist models with
more general charges.
We have seen in section \ref{ss:beq6}
that a rational section can intersect fibral rational curves
with an arbitrary intersection number.
Therefore it is sensible to expect that there exist
six-dimensional supergravity theories
with more general charges --- charges greater than
$2$ --- that are embeddable in F-theory.

An efficient strategy of finding F-theory backgrounds
with hypermultiplets of charge $\geq 3$ might be to
first construct $SU(2)$ theories with hypermultiplets
in higher-spin representations in addition to adjoints,
and then to obtain the $U(1)$ theories by Higgsing its
adjoint hypermultiplets.
For example, if there exists an
$SU(2)$ theory with a hypermultiplet in the
representation $J=3/2$ along with an adjoint ($J=1$,)
one can obtain a $U(1)$ theory
with hypermultiplets of charge $\pm 3$
by Higgsing the adjoint field.
It would be interesting to see if one could find
all such $SU(2)$ theories at least in the case
when $T=0$. There is reason to be optimistic
about this goal, given recent developments
on the space of $T=0$ theories
such as \cite{KPT,VBraun}.

A question that follows is whether there exist
$U(1)$ models in F-theory that cannot be enhanced
to $SU(2)$. We are not aware of any reason to
believe that such models do not exist.
If such models exist, however,
engineering them is expected to
be an algebraic challenge for reasons that
could be deduced from the way we have
constructed elliptic fibrations with Mordell-Weil
rank-one in appendix \ref{ap:2sect}.
Let us present the argument restricting to
the case when $T=0$ for sake of simplicity.

It is shown in appendix \ref{ap:2sect}
that the Mordell-Weil generator
of a threefold can be written in the form
\be
\sh~:\quad [x,y,z]
=[c_3^2-{2 \ov 3}b^2 c_2, -c_3^3+b^2 c_2 c_3 -{1 \ov 2} b^4 c_1,b] \,.
\ee
in Weierstrass coordinates.
If there are enough degrees of freedom in $b$ for it
to be tuned to $0$, this model can be enhanced
to an $SU(2)$ theory, as described at the end of
appendix \ref{ap:2sect}.
This is indeed the case for all the models we have
constructed in this note --- in fact, $b$ is an
arbitrary polynomial of degree $n$, whose coefficients
could all be tuned to zero for the manifolds
presented in section \ref{s:rank1}.
Therefore, in order for its low-energy $U(1)$ theory
to be ``un-enhancable," an elliptically fibered threefold
with Mordell-Weil rank-one must be ``rigid,"
in the sense that its complex structure must be fixed
at a certain point.
Such loci in the F-theory moduli space are
difficult to find.

It would nevertheless be interesting to identify
such models and compare them to what is allowed
from anomaly constraints. 
If the string universality conjecture \cite{KTUniv}
holds, there should be a correspondence between
non-trivial solutions of the $U(1)$ anomaly equations
\eq{anom1}, \eq{anom2} and these special un-enhancable
points in the F-theory moduli space.
Whether such a correspondence indeed
exists remains to be seen.
\vspace{0.1in}

\noindent
{\bf A Generalized Kodaira Constraint} \\
We return to the question that initiated our study
of models with Mordell-Weil rank-one --- is there
a generalized version of the Kodaira constraint
for the abelian sector of six-dimensional
F-theory models?
In the case of $T=0$ F-theory models with
gauge group $U(1)$, we have simplified the question.
Recall that these models come from compactifying F-theory on
elliptically fibered Calabi-Yau threefolds over $\field{P}^2$
with Mordell-Weil rank-one and no fibral divisors.
The \Neron-Tate height of a rational section
of such a manifold is given by a number.
The analogue of the Kodaira condition in this case
would be a bound on the \Neron-Tate height of the
generator of the Mordell-Weil group.
From arguments presented in the introduction,
such a bound should indeed exist ---
it would be very interesting to find what
that bound is.

In the event that such a bound on the height of the
Mordell-Weil generator is attained,
it would be interesting to see how it is modified
in more general situations.
For example, this bound might be modified
when there is non-abelian gauge symmetry.
Also, when the Mordell-Weil rank is larger than one
--- {\it i.e.,} when there are multiple $U(1)$'s ---
such a bound is expected to generalize to
a constraint on the height-pairing matrix of
the basis of the Mordell-Weil group.
We can further generalize to theories with $T>0$
--- {\it i.e.,} when the base of the elliptic
fibration is a general rational surface
rather than a $\field{P}^2$ \cite{MTBases1,MTBases2,TaylorHodge}.
In this case, the height pairing of rational sections become
divisors in the base, rather than numbers.
Such bounds, if attained, will play a crucial role in
gaining a better understanding of the space of
six-dimensional F-theory vacua, and ultimately
the space of six-dimensional supergravity theories.

\section*{Acknowledgement}

We thank Volker Braun, Mboyo Esole,
Antonella Grassi, Thomas Grimm,
Christoph Mayrhofer, Sug Woo Shin
and Wati Taylor for useful discussions.
D.P. would especially like to thank Wati Taylor for
his support and encouragement throughout the
course of this work.
We would also like to thank the Simons Center for Geometry
and Physics
and the organizers of the 2012 Summer Simons Workshop
in Mathematics and Physics
for their hospitality while part of this work was carried out.
In addition, D.R.M. thanks the Aspen Center for Physics 
and D.P. thanks
the organizers of String Phenomenology 2012
for hospitality.
This work is supported in part by funds provided by the
DOE under contract \#DE-FC02-94ER40818 and by National Science Foundation
grants DMS-1007414 and PHY-1066293.
D.P. also acknowledges support as a
String Vacuum Project Graduate Fellow,
funded through NSF grant PHY/0917807.

\appendix

\section{Addition of Sections} \label{ap:add}

An elliptic curve over a field may be written in Weierstrass
form:
\begin{align}
\begin{split}
y^2 &= x^3 + fxz^4 +gz^6
\end{split}
\end{align}
as a hypersurface of $\field{P}[2,3,1]$,
where $(x,y,z)$ are its projective coordinates.

Let us define the ``zero point" of the elliptic curve to be at
$(x,y,z)=(1,1,0)$. Working in the affine chart $z=1$,
we can now define the addition ``$\pls$" of two points
$p=(a,b)$ and $P=(A,B)$ on an elliptic curve over a field.
The symbol ``$\pls$" is used for this algebraic
addition to distinguish
from addition defined in the homology ring.
Note that
\begin{align}
\begin{split}
y^2 &= x^3 + fx +g \\
&= (x-a)(x^2+ax+c)+b^2 \\
&=(x-A)(x^2+Ax+C)+B^2 \,. \\
\end{split}
\end{align}

The new point $\mathcal{P}=p\pls P=(\mathcal{A},\mathcal{B})$
is obtained by demanding that $(\mathcal{A},-\mathcal{B})$ is
the third intersection point of the line
that goes through the two points $p$ and $P$.
It can easily be shown that
\be
\mathcal{P} = \left( \left( {B-b \ov A-a} \right)^2 -(a+A),
-\left( {B-b \ov A-a} \right)^3+(2a+A) \left( {B-b \ov A-a} \right)-b
\right) \,.
\ee

One could also find $\mathcal{P}'=P\pls P=(\mathcal{A}',\mathcal{B}')$
by demanding that $(\mathcal{A}',-\mathcal{B}')$ is
the other intersection point of the tangent line of the elliptic curve
that goes through $P$.
$\mathcal{P}'$ is given by
\be
\mathcal{P}' = \left( \left( {C+2A^2 \ov 2B} \right)^2 -2A,
-\left( {C+2A^2 \ov 2B} \right)^3+3A \left( {C+2A^2 \ov 2B} \right)-B
\right) \,.
\ee

It can be shown that the rational points of the elliptic curve
form an abelian group under the group action ``$\pls$" ---
the action is commutative and associative, and the zero
point is the identity element of the action.
It is also clear that $a\pls b$ is a rational point
when $a$ and $b$ are rational points.

\section{Elliptic Fibrations with Two Sections} \label{ap:2sect}

In this appendix, we construct the Weierstrass
model for elliptic fibrations with two rational sections
over a field $K$. We begin by reviewing how to arrive
at a Weierstrass model given the condition that there
exists one section. We proceed to obtain the Weierstrass
model when there are two sections.

\begin{figure}[!t]

\setlength{\unitlength}{.8cm} 

\noindent
\hfill
\begin{picture}(2,2)
\put(0,2){Variety}
\end{picture}
\hfill
\begin{picture}(4,2)
\put(2,2){$M$}
\end{picture} 
\hfill
\begin{picture}(4,2)
\put(2,2){$N$}
\end{picture} 
\hfill
\quad

\vskip-48pt

\noindent
\rule{\textwidth}{0.4mm}

\noindent
\hfill
\begin{picture}(2,4)
\put(0,2){$\mathbb{P}^{(1,2,3)}$}
\end{picture}
\hfill
\begin{picture}(4,4)
\thicklines
\put(-0.2,3.2){$z^6$}
\put(2.9,3.2){$x^3$}
\put(1.9,0.5){$y^2$}
\put(0,3){\line(1,0){3}}
\put(3,3){\line(-1,-2){1}}
\put(2,1){\line(-1,1){2}}
\put(0,3){\circle*{.2}}
\put(1,2){\circle*{.2}}
\put(1,3){\circle*{.2}}
\put(2,1){\circle*{.2}}
\put(2,3){\circle*{.2}}
\put(3,3){\circle*{.2}}
\put(2,2){\circle{.2}}
\end{picture} 
\hfill
\begin{picture}(4,4)
\thicklines
\put(-0.2,3.2){$D_z$}
\put(2.9,3.2){$D_x$}
\put(2.0,0.5){$D_y$}
\put(0,3){\line(1,0){3}}
\put(3,3){\line(-1,-2){1}}
\put(2,1){\line(-1,1){2}}
\put(0,3){\circle*{.2}}
\put(1,2){\circle*{.2}}
\put(1,3){\circle*{.2}}
\put(2,1){\circle*{.2}}
\put(2,3){\circle*{.2}}
\put(3,3){\circle*{.2}}
\put(2,2){\circle{.2}}
\end{picture} 
\hfill
\quad

\noindent
\hfill
\begin{picture}(2,4)
\put(0,2){$\mathbb{P}^{(1,1,2)}$}
\end{picture}
\hfill
\begin{picture}(4,4)
\thicklines
\put(-0.2,3.2){$u^4$}
\put(3.9,3.2){$v^4$}
\put(1.9,0.5){$w^2$}
\put(0,3){\line(1,0){4}}
\put(4,3){\line(-1,-1){2}}
\put(2,1){\line(-1,1){2}}
\put(0,3){\circle*{.2}}
\put(1,2){\circle*{.2}}
\put(1,3){\circle*{.2}}
\put(2,1){\circle*{.2}}
\put(2,3){\circle*{.2}}
\put(3,2){\circle*{.2}}
\put(3,3){\circle*{.2}}
\put(4,3){\circle*{.2}}
\put(2,2){\circle{.2}}
\end{picture} 
\hfill
\begin{picture}(4,4)
\thicklines
\put(0.2,3.2){$D_u$}
\put(2.9,3.2){$D_v$}
\put(2.0,0.5){$D_w$}
\put(1,3){\line(1,0){2}}
\put(3,3){\line(-1,-2){1}}
\put(2,1){\line(-1,2){1}}
\put(1,3){\circle*{.2}}
\put(2,1){\circle*{.2}}
\put(2,3){\circle*{.2}}
\put(3,3){\circle*{.2}}
\put(2,2){\circle{.2}}
\end{picture} 
\hfill
\quad

\noindent
\hfill
\begin{picture}(2,4)
\put(0,2){$\operatorname{Bl}_{[0,1,0]}\mathbb{P}^{(1,1,2)}$}
\end{picture}
\hfill
\begin{picture}(4,4)
\thicklines
\put(-0.2,3.2){$u^4$}
\put(2.9,3.2){$uv^3$}
\put(3.2,2.0){$v^2w$}
\put(1.9,0.5){$w^2$}
\put(0,3){\line(1,0){3}}
\put(3,3){\line(0,-1){1}}
\put(3,2){\line(-1,-1){1}}
\put(2,1){\line(-1,1){2}}
\put(0,3){\circle*{.2}}
\put(1,2){\circle*{.2}}
\put(1,3){\circle*{.2}}
\put(2,1){\circle*{.2}}
\put(2,3){\circle*{.2}}
\put(3,2){\circle*{.2}}
\put(3,3){\circle*{.2}}
\put(2,2){\circle{.2}}
\end{picture} 
\hfill
\begin{picture}(4,4)
\thicklines
\put(0.2,3.2){$D_u$}
\put(2.9,3.2){$D_v$}
\put(2.0,0.5){$D_w$}
\put(0.2,2.0){$E_1$}
\put(1,3){\line(1,0){2}}
\put(3,3){\line(-1,-2){1}}
\put(2,1){\line(-1,1){1}}
\put(1,2){\line(0,1){1}}
\put(1,2){\circle*{.2}}
\put(1,3){\circle*{.2}}
\put(2,1){\circle*{.2}}
\put(2,3){\circle*{.2}}
\put(3,3){\circle*{.2}}
\put(2,2){\circle{.2}}
\end{picture} 
\hfill
\quad

\noindent
\hfill
\begin{picture}(2,4)
\put(0,2){$Z$}
\end{picture}
\hfill
\begin{picture}(4,4)
\thicklines
\put(0.3,3.2){$u^3v$}
\put(2.9,3.2){$uv^3$}
\put(3.2,2.0){$v^2w$}
\put(1.9,0.5){$w^2$}
\put(1,3){\line(1,0){2}}
\put(3,3){\line(0,-1){1}}
\put(3,2){\line(-1,-1){1}}
\put(2,1){\line(-1,2){1}}
\put(1,3){\circle*{.2}}
\put(2,1){\circle*{.2}}
\put(2,3){\circle*{.2}}
\put(3,2){\circle*{.2}}
\put(3,3){\circle*{.2}}
\put(2,2){\circle{.2}}
\end{picture} 
\hfill
\begin{picture}(4,4)
\thicklines
\put(0.2,3.2){$D_u$}
\put(2.9,3.2){$D_v$}
\put(3.9,3.2){$E_3$}
\put(2.0,0.5){$D_w$}
\put(0.2,2.0){$E_1$}
\put(3.2,1.8){$E_2$}
\put(1,3){\line(1,0){3}}
\put(4,3){\line(-1,-1){2}}
\put(2,1){\line(-1,1){1}}
\put(1,2){\line(0,1){1}}
\put(1,2){\circle*{.2}}
\put(1,3){\circle*{.2}}
\put(2,1){\circle*{.2}}
\put(2,3){\circle*{.2}}
\put(3,2){\circle*{.2}}
\put(3,3){\circle*{.2}}
\put(4,3){\circle*{.2}}
\put(2,2){\circle{.2}}
\end{picture} 
\hfill
\quad

\caption{\small Toric data for ambient spaces of some elliptic curve
embeddings.  These are among the 16 reflexive toric surfaces 
\cite{MR1463052} (see also \cite{arXiv:1201.0930}).
\label{f:toric}}
\end{figure}
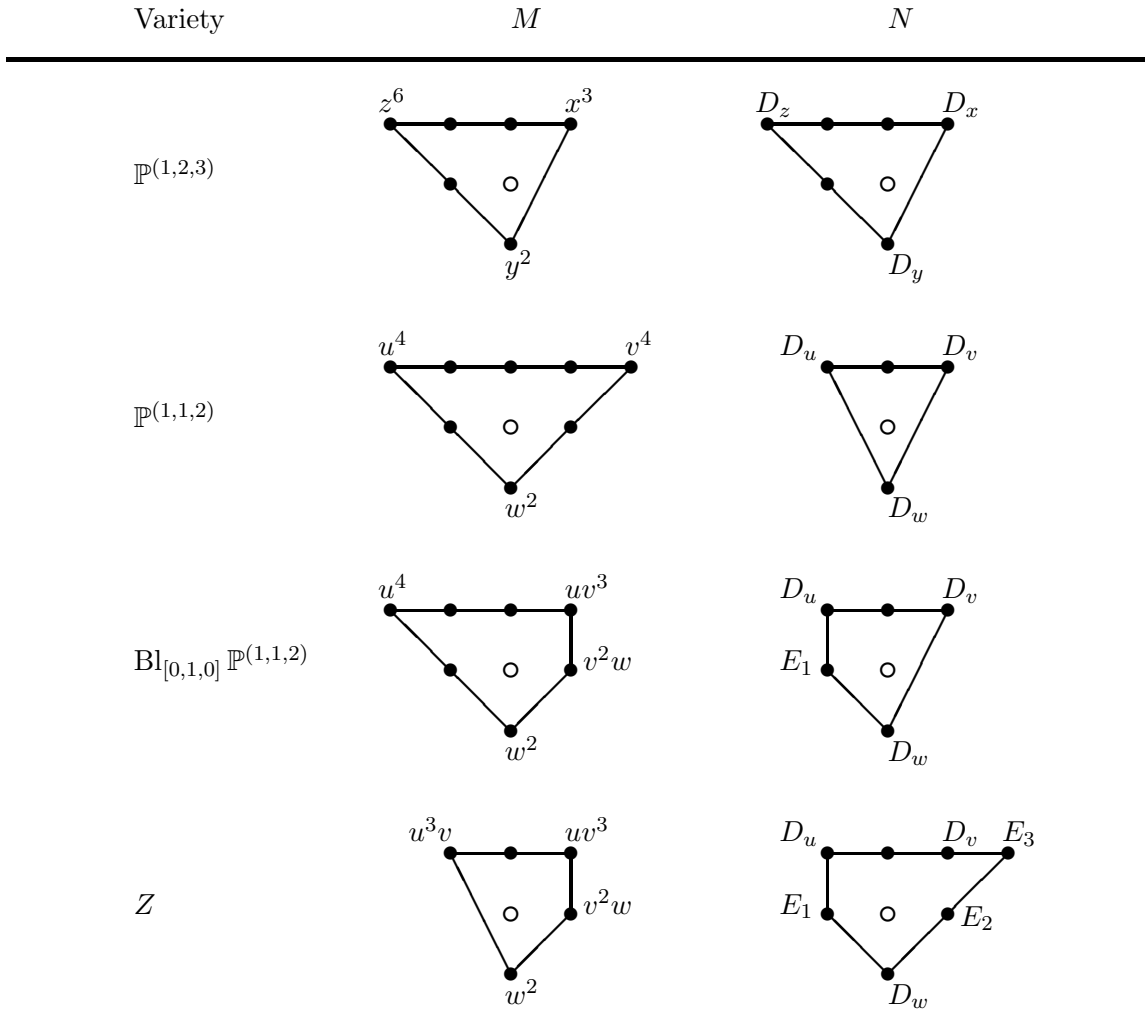

Let us first review how to arrive at the Weierstrass model
of an elliptic curve $E$ over a field $K$ with a point $P$,
or more generally, over a ring $R$ whose fraction field is
$K$.\footnote{In the context of this note, $R$ is the coordinate
ring of the base manifold, $K$ is the function field of the base,
and $P$ is the ``zero section" of the elliptic fibration.}
We start with the line bundle $L = \OO(P)$
and consider sections: $H^0(L)$
has a single section, denoted by $z$.
$H^0(2L)$ has two sections,
one of which is $z^2$ and the other of which is new,
which we denote $x$. $H^0(3L)$ has three sections:
$z^3$, $xz$, and a new one $y$.
$H^0(4L)$ has four sections: $z^4$, $xz^2$,
$yz$, and $x^2$. $H^0(5L)$ has five sections:
$z^5$, $xz^3$, $yz^2$, $x^2z$, and $xy$.
$H^0(6L)$ should only have six sections,
but we know about seven:
$z^6$, $xz^4$, $yz^3$, $x^2z^2$, $xyz$, $x^3$, $y^2$.
Thus, there must be a relation,
and one argues --- following Deligne \cite{Deligne} ---
that the coefficients of $x^3$ and $y^2$
must be units in the ring $R$ and after an
appropriate scaling, we get a Weierstrass equation of the form
\be
y^2 + a_1xyz + a_3 yz^3 = x^3 + a_2x^2z^2 + a_4xz^4 + a_6z^6 \,.
\label{firstW}
\ee
Since the variables $z$, $x$, and $y$ have weights $1$, $2$, and $3$,
this can be regarded as a hypersurface in the weighted projective
space $\mathbb{P}^{(1,2,3)}$, which as a toric variety is illustrated
in the first row of figure \ref{f:toric}.
The monomials which occur in equation \eq{firstW} are indicated
as a polytope contained in the monomial lattice $M$, and the toric divisors
$D_x$, $D_y$, and $D_z$ are indicated as the generators of the
polar polytope in the dual lattice $N$.

Typically the Weierstrass equation is studied in the affine chart $z = 1$.
If the characteristic of $K$ is not 2 or 3 (which is true in our case),
we can complete the square in $y$ and then complete
the cube in $x$, resulting in an equation with  $a_1 =a_2 =a_3 =0$.

Now let us consider the case with two sections.
Suppose that we have an elliptic curve over a field
$K$ with two points $P$ and $Q$,
coming from two sections of the elliptic fibration.
We assume that both points are defined over $K$,
but do not assume that they are necessarily distinct.
This time we use the line bundle $M = \OO(P + Q)$
and again study sections.
We first study sections and an embedding in the case of an arbitrary
line bundle $M$ of degree $2$, and subsequently specialize to the case 
of $M = \OO(P + Q)$.

Since $H^0(M)$ has two sections,  we let $u$, $v$ be a basis of this space.
Now $H^0(2M)$ has four sections:
$u^2$, $uv$, $v^2$, and a new one which we denote by $w$.
The space $H^0(3M)$ has six sections, all of which are known:
$u^3$, $u^2 v$, $uv^2$, $v^3$, $uw$ and $vw$.
Finally, $H^0(4M)$ should have eight sections,
but we know nine: $u^4$, $u^3v$, $u^2v^2$, $uv^3$,
$v^4$, $uvw$, $u^2 w $, $v^2w$, and $w^2$.
Thus, there must be an equation.
It is not hard to argue that the coefficient of $w^2$ 
must be a unit (in order that
the solution set be a genus one curve)
and that by scaling we can set that
coefficient equal to 1:\footnote{Note that this equation is somewhat
more general than the ``$E_7$-fibrations'' studied in
\cite{Aldazabal:1996du,Berglund:1998va,Aluffi:2009tm}.
$E_7$-fibrations have been utilized
to obtain F-theory models with
abelian gauge symmetry
in the string phenomenology literature,
for example, in \cite{GrimmWeigand}.}
\be
w^2+b_0 u^2w +b_1uvw+b_2 v^2 w = c_0 u^4+c_1u^3v+c_2u^2v^2+c_3u v^3+c_4v^4 \,.
\label{deg2}
\ee
Since the variables $u$, $v$, and $w$ have weights $1$, $1$, and $2$, we can
regard this as defining a hypersurface in the weighted projective
space $\mathbb{P}^{(1,1,2)}$, which as a toric variety is illustrated
in the second row of figure \ref{f:toric}.
Again, the monomials in \eq{deg2} are shown, as well as the toric
divisors $D_u$, $D_v$, and $D_w$.

We can specialize the form of the equation further if we assume that
$M = \OO(P + Q)$.  In this case, we choose $u$ to be a section which
vanishes precisely at $P$ and $Q$, and let $v$ be an arbitrary second
section which vanishes elsewhere.
When we set $u = 0$ in \eq{deg2}, we get
\be
w^2 + b_0v^2w = c_4v^4
\ee
and so the two roots of this equation must
correspond to $P$ and $Q$.
Since $P$ and $Q$ are defined over the ground field $K$,
this equation must factor; then, shifting $w$ by an appropriate
multiple of $v$ we can assume that one of the factors is $w$,
{\it i.e.,} that $c_4 = 0$. 
This leaves us with an equation of the form 
\be
w^2+b_0 u^2w +b_1uvw+b_2 v^2 w = u(c_0u^3+c_1u^2v+c_2u v^2+c_3v^3) \,.
\label{deg2bis}
\ee
This is again a general hypersurface in a toric variety.
The difference between \eq{deg2} and \eq{deg2bis} is
that the monomial $v^4$ has been eliminated, as illustrated
in the third row of figure \ref{f:toric}.  The corresponding change
to the polar polytope corresponds to blowing up $\mathbb{P}^{(1,1,2)}$
at the point $[0,1,0]$, giving a new exceptional divisor which is
denoted by $E_1$.

Assuming the characteristic of $K$ is not 2,
we can further shift $w$ by a multiple of $u$ to also assume
that $b_0 = b_1 = 0$.
Let us simplify notation and denote $b_2$ simply by $b$.
Thus we obtain an equation of the form
\be
w^2 + bv^2w = u(c_0u^3 + c_1u^2v + c_2 u v^2 + c_3 v^3) \,,
\label{twosections}
\ee
in which the point $P$ has $[u,v,w] = [0,1,0]$ and the point $Q$
has $[u,v,w] = [0,1,-b]$.

Let us find the Weierstrass form of this fibration \eq{twosections}
corresponding to the section $P$.
For this purpose, we need to find sections of $H^0(kM-kQ)$,
that is, sections of $H^0(kM)$ which vanish $k$ times along $Q$.
The first of these is easy:  the section $u$ vanishes along Q,
and so in our construction of a Weierstrass model we take
\be
z := u \,.
\ee
Now we need sections of $H^0(2M)$,
that is, linear combinations of $w$ and a quadratic in $v$ and $u$.
One such section is $u^2$;
to find another, we set
\be
w = \alpha v^2 + \beta uv + \gamma u^2
\ee
and substitute in the equation:
\begin{align}
\begin{split}
&u(c_0u^3+c_1u^2v+c_2u v^2+c_3v^3) \\
&= w (w +bv^2) = (\alpha v^2+\beta uv+\gamma u^2)
((b+\alpha)v^2+\beta uv+\gamma u^2) \,.
\end{split}
\end{align}
To get $Q$ at $u=0$ we need $\alpha=-b$.
Thus, our equation becomes
\be
u(c_0 u^3 +c_1 u^2 v +c_2u v^2+c_3v^3)
= (-b v^2+\beta uv+ \gamma u^2)(\beta uv+ \gamma u^2) \,.
\ee
We need a double zero at $u=0$, which requires $c_3 = -b\beta$.
Thus, we should take $\beta = -c_3/b$ and hence
\be
w = -bv^2 -(c_3/b)uv + \gamma u^2 \,.
\ee
We can omit the $u^2$ term since
it is another solution.

More generally, we can clear denominators,
and take the second element of our Weierstrass form to be
\be
x := b^2 v^2 + bw + c_3uv \,.
\ee
Next, we need sections of $H^0(3M)$ vanishing three times
at $Q$. Two of these are $u^3$ and
$u(b^2v^2 + bw + c_3uv)$, so we seek a section of the form
$vw = \alpha v^3 + \beta uv^2 + \gamma u^2v$
omitting terms of the form $uw$ and $u^3$
since they are taken care of by other sections.
In this case, we substitute into the equation as follows:
\begin{align}
\begin{split}
&uv^2(c_0u^3+c_1u^2v+c_2uv^2+c_3v^3) \\
&= (vw)(vw +bv^3)
= (\alpha v^3+\beta uv^2+\gamma u^2v)((b+\alpha)v^3+\beta uv^2+\gamma u^2v) \,.
\end{split}
\end{align}
As in the previous case, we need $\alpha =-b$
to guarantee that at $u = 0$ we are getting $Q$,
which leads to an equation
\be
uv^2(c_0u^3+c_1 u^2v +c_2 u v^2+c_3 v^3)
= (-bv^3+\beta uv^2+\gamma u^2v)(\beta uv^2+\gamma u^2v) \,.
\ee
To get a triple zero at $u = 0$, we then require
\begin{align}
\begin{split}
c_2 &=-b\gamma + \beta^2\\
c_3 &=-b\beta 
\end{split}
\end{align}
which is solved by
\begin{align}
\begin{split}
\alpha&=-b\\
\beta&=-c_3 /b \\
\gamma&=-c_2/b +c_3^2/b^3 \,.
\end{split}
\end{align}

Thus, to complete our mapping to a Weierstrass model,
we mostly clear denominators and use
\be
y :=  b^2vw + b^3v^3 + bc_3uv^2 + (bc_2 -{c_3^2 \ov b})u^2v \,.
\ee
The Weierstrass equation in these coordinates is given by
\be
y^2-x^3+\frac{2c_3}bxyz+{c^2_3 -b^2c_2 \ov b^2} x^2z^2+bc_1yz^3+b^2c_0xz^4+c_0(b^2c_2-c^2_3)z^6 =0 \,.
\ee
By the reparametrization
\begin{align}
\begin{split}
\tx &= x+{c_2 \ov 3} z^2
= b^2 v^2 +c_3 uv + {c_2 \ov 3} u^2 + bw\\
\ty &= y+{c_3 \ov b} xz + {bc_1 \ov 2} z^3
= b^3 v^3 +2b c_3 uv^2 + b c_2 u^2 v + { b c_1 \ov 2} u^3 +c_3 uw+b^2 vw\\
z &= u
\end{split}
\label{e7toe8}
\end{align}
we arrive at the standard Weierstrass form:
\be
\ty^2=\tx^3 + (c_1 c_3 -b^2 c_0 - {c_2^2 \ov 3} )\tx z^4
+\left( c_0c_3^2  -{1 \ov 3} c_1c_2c_3 +{2 \ov 27} c_2^3
-{2 \ov 3} b^2 c_0c_2 +{b^2 c_1^2 \ov 4} \right) z^6 \,.
\label{ws}
\ee

To find the section $Q$ explicitly, we return to sections of
$H^0(2M)$, this time looking for a section which vanishes
three times at $Q$.
In our setup above, we have used $w = \alpha v^2 + \beta uv + \gamma u^2$,
and found the condition to vanish to order-two at $Q$.
Now we need to vanish to order-three,
which gives one additional equation:
\be
c_2 = -b\gamma + \beta^2 \,.
\ee
The solution, after normalization, is
\be
b^2v^2 +bw +c_3uv+(c_2 -{c^2_3 \ov b^2} )u^2 =x+(c_2 -{c^2_3 \ov b^2} )z^2 \,.
\ee
In other words, the $x$-coordinate of $Q$ is $((c_3/b)^2 -c_2)z^2$.
Substituting the corresponding $\tx$ value into equation \eq{ws},
we can solve for $\ty$.
The section can in fact be located at
\be
[\tx,\ty,z]=[c_3^2-{2 \ov 3}b^2 c_2, -c_3^3+b^2 c_2 c_3 -{1 \ov 2} b^4 c_1,b] \,.
\label{rank1sect}
\ee

The transition from an extra section to an enhanced $SU(2)$
is obtained when $b$ becomes identically zero,
which means that the sections $P$ and $Q$ are exactly the same.
It can be seen that the fiber at $c_3 = 0$ has Kodaira type $I_2$,
{\it i.e.,} it is an $SU(2)$ fiber.

\section{A Degenerate Limit}
\label{ap:C}

Suppose we have an elliptic fibration with two sections
for which the coefficients $b_0$ and $c_0$
in \eq{deg2bis} vanish identically, or equivalently, after completing
the square, the coefficent $c_0$ in 
\eq{twosections} vanishes identically.  In this case, the
ambient toric variety changes dramatically, as indicated in 
the fourth row of figure \ref{f:toric}.  

First, setting $c_0=0$ in \eq{deg2bis} corresponds to a second blowup
of $\mathbb{P}^{(1,1,2)}$ at $[1,0,0]$, giving an exceptional divisor
$E_2$.  Then, setting $b_0=0$ in \eq{deg2bis} corresponds to a third
blowup with corresponding exceptional divisor $E_3$.  At this
stage, however, the divisors $D_v$ and $E_2$ have intersection number zero with
the canonical divisor of the toric surface, so they are blown down
in the anti-canonical model of the toric variety.

More significant than the change in the toric variety, however, is
the behavior of the discriminant locus of the Weierstrass equation.
The Weierstrass equation for this family is determined by setting
$c_0=0$ in \eq{ws} (since we have already completed the square),
yielding
\be
\ty^2=\tx^3 + (c_1 c_3 - {c_2^2 \ov 3} )\tx z^4
+\left( -{1 \ov 3} c_1c_2c_3 +{2 \ov 27} c_2^3
+{b^2 c_1^2 \ov 4} \right) z^6 \,.
\label{wsdegenerate}
\ee
It is straightforward to compute the discriminant of \eq{wsdegenerate}
and we find:
\be
\frac1{16}c_1^2
\left(27b^4c_1^2+16b^2c_2^3-72b^2c_1c_2c_3-16c_2^2c_3^2+64c_1c_3^3\right)\,.
\ee
Note that when $c_1=0$, the coefficients of $\tx z^4$ and $z^6$ do
not necessarily vanish.

The interpretation of the factor of $c_1^2$ in this discriminant is
as follows.  Whenever we have an elliptic fibration with two sections
of this form --- with the coefficients being sections of appropriate
line bundles over the base --- the fibration will have
fibers of Kodaira type $I_2$ along the locus $c_1 =0$.  In particular,
in F-theory there will be a locus with enhanced $SU(2)$ gauge symmetry.

The candidate models for $\sT_7$ and $\sT_8$ discussed
in section \ref{ss:more} are precisely of this form ---
these manifolds were constructed by setting $f_{12-2n}=0$
in the ansatz \eq{rank1ansatz}.
We now see that those constructions do not have simply a $U(1)$
gauge symmetry, but have an additional $SU(2)$ gauge symmetry,
which is not what is desired.

Note that there is one exception to this conclusion, that is, when
$c_1$ itself is nowhere-vanishing.  In the body of the paper, we have
considered a situation in which $c_1$ is a polynomial
of degree $(9-n)$ on $\mathbb{P}^2$.
If $n=9$, $c_1$ does not vanish and indeed the ``$\sT_9$ theory''
agrees with the $\sT_0$ theory but with the ``wrong'' choice of generating
section.

\end{document}